\newif\ifpreprint

\preprintfalse 

\ifpreprint
\documentclass[aip,jcp,amsmath,amssymb,preprint]{revtex4-1}
\else
\documentclass[aip,jcp,amsmath,amssymb,reprint]{revtex4-1}
\fi

\usepackage{xspace}
\usepackage{graphicx}
\usepackage{bm}
\usepackage{amsfonts}
\usepackage[version=4]{mhchem}
\usepackage{pbox}
\usepackage{physics}
\usepackage{multirow}
\usepackage{threeparttable}
\usepackage{float}
\usepackage{xspace}
\usepackage{color}
\usepackage{fancyvrb}
\usepackage{adjustbox}
\usepackage{tikz}
\usepackage{marvosym}
\usetikzlibrary{shapes.geometric, arrows}
\usepackage[note-name=,use-sort-key=false]{notes2bib}
\usepackage[colorlinks = true,
            linkcolor = blue,
            urlcolor  = black,
            citecolor = blue,
            anchorcolor = black]{hyperref}
\usepackage{textcase}

\newcommand*{\sunit}{$E_{\rm h}^{-2}$\xspace}
\newcommand*{\Eh}{$E_{\rm h}$\xspace}

\newcommand*{\PSI}{{\scshape Psi4}\xspace}

\newcommand*{\forte}{{\scshape Forte}\xspace}

\newcommand{\mref}[0]{\Psi_0}
\newcommand{\tens}[3]{{#1}_{#2}^{#3}}
\newcommand{\dfock}[1]{\epsilon_{#1}}
\newcommand{\cop}[1]{\hat{a}^\dag_{#1}}
\newcommand{\aop}[1]{\hat{a}_{#1}}
\newcommand{\sqop}[2]{\hat{a}_{#2}^{#1}}
\newcommand{\nsqop}[2]{\{\hat{a}_{#2}^{#1}\}}
\newcommand{\qop}[1]{\hat{q}_{#1}}

\newcommand{\no}[1]{ \{ {#1} \}}

\newcommand{\permop}{{\cal \hat{P}}}
\newcommand{\vtau}{{\bm{\tau}}}

\usepackage{newtxtext,newtxmath}
\usepackage[scaled=1.0]{helvet}
\usepackage[compact]{titlesec}
\usepackage{xcolor}
\usepackage{notes2bib}
\usepackage{chemformula}
\bibnotesetup{ note-name = , use-sort-key = false}
\titleformat{\section}
  {\normalfont\sffamily\bfseries}
  {\thesection.}{0.5 em}{\MakeTextUppercase}
\titlespacing{\section}{0pt}{12pt}{12pt}

\titleformat{\subsection}[block]
  {\normalfont\sffamily\bfseries}
  {\thesubsection.}{0.5 em}{}
\titlespacing{\subsection}{0pt}{12pt}{8pt}

\titleformat{\subsubsection}[block]
  {\normalfont\itshape\sffamily\bfseries\raggedright}
  {\arabic{subsubsection}.}{0.5 em}{}
\titlespacing{\subsubsection}{0pt}{8pt}{8pt}

\usepackage[normalem]{ulem}

\definecolor{goodorange}{RGB}{225,125,0}
\definecolor{goodgreen}{RGB}{0,125,0}
\definecolor{goodred}{RGB}{220,50,25}
\definecolor{goodblue}{RGB}{25,25,150}

\newcommand{\note}[2]{
\ifthenelse{\equal{#1}{F}}{
\colorbox{goodorange}{\textcolor{white}{\footnotesize \fontfamily{phv}\selectfont #1}}
    \textcolor{goodorange}{{\footnotesize \fontfamily{phv}\selectfont #2}}\xspace
}{}
\ifthenelse{\equal{#1}{Y}}{
\colorbox{goodred}{\textcolor{white}{\footnotesize \fontfamily{phv}\selectfont #1}}
    \textcolor{goodred}{{\footnotesize \fontfamily{phv}\selectfont #2}}\xspace
}{}
}

\usepackage{dcolumn}
\newcolumntype{d}[1]{D{.}{.}{#1}}

\newcommand\trick[1]{}

\begin{document}

\title{Connected three-body terms in single-reference unitary many-body theories: Iterative and perturbative approximations}

\author{Chenyang Li}
\email{bjyork0822@gmail.com}
\affiliation{Department of Chemistry and Cherry Emerson Center for Scientific Computation, Emory University, Atlanta, GA 30322, USA}

\author{Francesco A. Evangelista}
\email{francesco.evangelista@emory.edu}
\affiliation{Department of Chemistry and Cherry Emerson Center for Scientific Computation, Emory University, Atlanta, GA 30322, USA}

\date{\today}

\begin{abstract}
This work introduces various approaches to include connected three-body terms in unitary many-body theories, focusing a representative example on the driven similarity renormalization group (DSRG).
Starting from the least approximate method---the linearized DSRG truncated to one-, two-, and three-body operators [LDSRG(3)]---we develop several approximate LDSRG(3) models with reduced computational cost.
Through a perturbative analysis, we motivate a family of iterative LDSRG(3)-$n$ and -$n'$ ($n=1,2,3,4$) methods that contain a subset of the LDSRG(3) diagrams.
Among these variants, the LDSRG(3)-2 scheme has the same computational complexity of coupled cluster theory with singles, doubles, and triples (CCSDT), but it outperforms CCSDT in the accuracy of the predicted correlation energies.
We also propose and implement two perturbative triples corrections based on the linearized DSRG truncated to one- and two-body operators augmented with recursive quadratic commutators [qDSRG(2)].
The resulting qDSRG(2)+(T) approach matches the accuracy of the ``gold-standard'' coupled cluster theory with singles, doubles, and perturbative triples model on the energetics of twenty-eight closed-shell atoms and small molecules.
\end{abstract}

\maketitle

\section{Introduction}
\label{sec:intro}

With recent advances to quantum computing, there has been a revival of interest in nonperturbative electronic-structure theories based on unitary transformations.\cite{Peruzzo:2014kc,OMalley:2016dc,Shen:2017cc,Hempel:2018ip,Barkoutsos:2018hm,Grimsley:2019ed,Evangelista:2019kz}
In these methods, the exact wave function $\Psi$ is obtained from a normalized reference wave function $\mref$ via a unitary transformation $\hat{U}$:
\begin{align}
\label{eq:wfn_unitary}
  \ket{\Psi} = \hat{U} \ket{\mref} = e^{\hat{A}} \ket{\mref}.
\end{align}
Here, the wave operator is written as an exponential of an anti-Hermitian operator $\hat{A} = \hat{T} - \hat{T}^\dag$, parametrized using the coupled cluster (CC) excitation operator $\hat{T}$.\cite{Cizek:1966cy,Crawford:2000by,Bartlett:2007kv}
Combining Eq.~\eqref{eq:wfn_unitary} and the electronic Sch{\"o}dinger equation, we obtain the following energy expression
\begin{align}
\label{eq:Eunitary}
E = \ev{e^{\hat{A}^\dagger} \hat{H} e^{\hat{A}}}{\mref} = \ev{e^{-\hat{A}} \hat{H} e^{\hat{A}}}{\mref},
\end{align}
where $\hat{H}$ indicates the bare Born--Oppenheimer Hamiltonian.
The unitary transformed Hamiltonian ($\bar{H}$) can be expressed in a connected form using the Baker--Campbell--Hausdorff (BCH) formula:
\begin{align}
\label{eq:BCH}
\bar{H} &\equiv e^{-\hat{A}} \hat{H} e^{\hat{A}} \notag\\
&= \hat{H} + [\hat{H}, \hat{A}] + \frac{1}{2!}[ [ \hat{H}, \hat{A}], \hat{A}] + \frac{1}{3!}[ [ [ \hat{H}, \hat{A}], \hat{A}], \hat{A}] +\cdots .
\end{align}
Equations \eqref{eq:Eunitary} and \eqref{eq:BCH} highlight the advantages of unitary formalisms: the energy is both variational and size extensive.
Moreover, the transformed Hamiltonian is guaranteed to be Hermitian, which is appealing when computing properties and formulating multireference theories.

However, Eq.~\eqref{eq:BCH} yields a non-terminating series that cannot be evaluated unless both the operator $\hat{A}$ and the BCH series are approximated.
As in CC theory, $\hat{A}$ may be approximated by truncating $\hat{T}$ to a given substitution level (typically 2--3 body), which is often sufficient to recover correlation effects with high accuracy.
However, any truncation of the BCH series [Eq.~\eqref{eq:BCH}] results in the energy not being strictly variational.
Nonetheless, a number of truncation schemes to the BCH series have been proposed.
Initial attempts in unitary CC (UCC) theory kept only a finite number of nested commutators based on a perturbative argument.\cite{Bartlett:1989iz,Watts:1989uo,Watts:1989ve,Kutzelnigg:1991iw}
Numerical results have shown that four to six terms are necessary to achieve sub-m\Eh accuracy compared to the numerical exact infinite series.\cite{Evangelista:2011jp}
An alternative truncation scheme has also been suggested by Taube and Bartlett where the theory is formulated to be exact for a given number of electrons.\cite{Taube:2006bi}

Another approach to approximating the BCH series is to limit the many-body character of the nested commutators.
To this end, Yanai and Chan introduced the linear truncation scheme,\cite{Yanai:2006gi,*Yanai:2007ix} where \textit{each} single commutator in Eq.~\eqref{eq:BCH} is truncated to contain at most $i$-body components:
\begin{align}
\label{eq:rsc}
[\,\cdot\, , \hat{A}] \approx \sum_{k=0}^{i} [\,\cdot\, , \hat{A}]_k \equiv [\,\cdot\, , \hat{A}]_{\{i\}}.
\end{align}
Here, we have introduced a compact notation to indicate the $k$-body component of a commutator ($[\,\cdot\, , \hat{A}]_k$) and for the sum of many-body components of a commutator up to order $i$ ($[\,\cdot\, , \hat{A}]_{\{i\}}$).
One of the simplest schemes in this family of approximations assumes that $\hat{A} \approx \hat{A}_1 + \hat{A}_2$ and $[\,\cdot\, , \hat{A}] \approx [\,\cdot\, , \hat{A}]_{\{2\}}$, which is referred as the L2SD approximation in the following.
Since both the bare Hamiltonian and each single commutator contain at most two-body operators, only expressions for terms resulting from $[\hat{O}, \hat{A}]_{\{2\}}$ need to be derived, assuming an arbitrary operator $\hat{O}$ of the form $\hat{O} = \hat{O}_1 + \hat{O}_2$.
Consequently, the nested commutators in Eq.~\eqref{eq:BCH} can be computed recursively and the cost of every step scales as ${\cal O}(N_{\bf O}^2 N_{\bf V}^2 N_{\bf G}^2)$, where $N_{\bf O}$, $N_{\bf V}$, and $N_{\bf G}$ indicate the number of occupied, virtual, and general orbitals, respectively.
This scaling is asymptotically identical to that of CC with singles and doubles (CCSD)\cite{PurvisIII:1982kx}, yet the actual computational cost is roughly that of CCSD times the number of nested commutators included in the BCH series.
Numerical results show that the L2SD approach tends to overestimate correlation energies and it is not appropriate for computations aiming for high accuracy.\cite{Evangelista:2014kt,Li:2016hb,*Li:2018dy,Li:2017bx,*Li:2018fn}
One of its main deficiency is the lack of connected triples excitations.

In the context of CC theory, multiple schemes have been proposed to address the effects of triple excitations, which can be classified into iterative and perturbative methods.
The most comprehensive approach is the CCSD with full triples (CCSDT) model.\cite{Noga:1987ea,Scuseria:1988by}
The CCSDT energy is correct through fourth order in perturbation theory, yet its high computational complexity [${\cal O}(N_{\bf O}^3 N_{\bf V}^5)$] practically limits its application only to small molecules.
Successful attempts have been made to reduce the computational cost by approximating the CCSDT equations.\cite{Lee:1984fq,*Lee:1998gm,Urban:1985bq,Koch:1997eu}
For example, the CCSDT-1 method of Bartlett and co-workers includes only linear $\hat{T}_3$ terms in the wave function, yielding an asymptotic scaling of ${\cal O}(N_{\bf O}^3 N_{\bf V}^4)$.\cite{Lee:1984fq,*Lee:1998gm,Urban:1985bq}
Nevertheless, the CCSDT-1 scheme tends to overestimate the effect of triples and the corresponding equations need to be solved iteratively.\cite{Urban:1985bq,He:2001co}

Methods that include triples perturbatively avoid the iterative procedure and perform only one ${\cal O}(N_{\bf O}^3 N_{\bf V}^4)$ step after the CCSD computation.
The most widely used method in this category is the CCSD(T) model.\cite{Raghavachari:1989gf,Stanton:1997dz}
The (T) correction adds several energy terms on top of the CCSD energy.
The most important one is already included in the [T] correction,\cite{Urban:1985bq} which considers fourth-order energy contributions due to triples using the CCSD converged $\hat{T}_2$ amplitudes.
Like CCSDT-1, the [T] correction usually exaggerates the correlation energy due to triple excitations.\cite{Urban:1985bq}
To counterbalance this overestimation, Raghavachari \textit{et al.}\ consider a fifth-order energy term assuming the use of a Hartree--Fock (HF) reference.\cite{Raghavachari:1989gf}
It was later realized that this term and an extra term (that vanishes in the HF case) appear at the fourth-order energy for non-HF references.\cite{Watts:1993cc}
Thus, these three terms define the (T) correction in the most general way.

Inspired by the success of CCSD(T), significant effort has been devoted to developing systematic procedures for adding perturbative corrections to the CCSD energy.
A seminal perspective on CCSD(T) was given by Stanton,\cite{Stanton:1997dz} who showed that the (T) correction may be derived by applying L{\"o}wdin's partitioning technique to the CCSD similarity transformed Hamiltonian.
This idea later led to the antisymmetric CCSD(T) method of Crawford and Stanton\cite{Crawford:1998cq} and the equivalent $\Lambda$CCSD(T) method coined by Kucharski and Bartlett.\cite{Kucharski:1998ie,Taube:2008ca}
In $\Lambda$CCSD(T), both the CCSD cluster and lambda amplitudes determine the triples correction, with the latter obtained by solving an additional set of iterative equations.
Later developments of non-iterative triples include the completely renormalized CCSD(T) of Piecuch and co-workers,\cite{Kowalski:2000ja,Piecuch:2004kh,Piecuch:2005cd} the CCSD(2) approach of Gwaltney \textit{et al.},\cite{Gwaltney:2000fj} the CCSD(2)$_T$ scheme of Hirata \textit{et al.},\cite{Hirata:2004fm} and the CCSD(T-$n$) methods of Eriksen \textit{et al}.\cite{Eriksen:2014jk,*Eriksen:2015bd}
We note that the original (T) correction is a special case of all these approaches.
Despite the fact that many well-established methods exist to add connected triples in CC theory, to the best of our knowledge, there are no studies that have extensively explored the same issue in unitary theories.

Over the past few years, we have developed numerically robust multireference (MR) theories based on the unitary driven similarity renormalization group (DSRG) approach.\cite{Evangelista:2014kt,Li:2016hb,Li:2019fu}
In DSRG, a flow parameter is employed to systematically regularize the divergences resulting from zeroth-order degeneracies between the reference wave function and its excited configurations.
When the flow parameter goes to infinity, the single-reference DSRG and UCC equations become equivalent.
One of the simple non-perturbative realizations of the DSRG employs the L2SD approximation, leading to the LDSRG(2) scheme developed for both single-reference (SR) and multireference theories.\cite{Evangelista:2014kt,Li:2016hb}
A small benchmark of single-bond dissociations shows that the MR-LDSRG(2) approach yields small absolute errors along the potential energy curves.\cite{Li:2016hb,Li:2017bx}
However, this high accuracy deteriorates considerably when breaking multiple bonds.\cite{Li:2017bx}
It is thus important to develop more accurate approximations and to go beyond the MR-LDSRG(2) method.
As an initial attempt towards this goal, we explore the possibilities of introducing connected triples in the SR-DSRG framework.

In this work, we consider connected triples from a ``top--down'' perspective.
We start by formulating a full-fledged LDSRG(3) theory, where the L3SDT truncation scheme is employed, that is, assuming $\hat{A} \approx \hat{A}_1 + \hat{A}_2 + \hat{A}_3$ and $[\,\cdot\, , \hat{A}] \approx [\,\cdot\, , \hat{A}]_{\{3\}}$.
Unfortunately, the recursive evaluation of $\bar{H}_3$ in LDSRG(3) scales as ${\cal O}(N_{\bf O}^3 N_{\bf V}^3 N_{\bf G}^3)$, a cost significantly higher than that of CCSDT.
We then propose iterative models, designated as LDSRG(3)-$n$ ($n=1,2,3,4$), obtained by trimming the LDSRG(3) equations based on a perturbative assumption.
The simplest of these models, the LDSRG(3)-1 scheme, has a computational complexity identical to that of CCSDT-1.
To further reduce the computational pre-factor and storage cost, we also consider the possibilities of truncating the BCH expansion of $\bar{H}_3$.
To formulate perturbative triples corrections, we first define a pseudo-quadratic DSRG scheme  [qDSRG(2)], which accounts for the missing fourth-order terms in the L2SD approximation. Next, we introduce (T) and [T]-like corrections by approximating the fourth-order terms of the DSRG $\Lambda$(T) energy functional.
Computing the qDSRG(2)+(T)/[T] energy has a cost that is the sum of the qDSRG(2) procedure [iterative, ${\cal O}(N_{\bf O}^2 N_{\bf V}^4)$ scaling] plus the evaluation of the perturbative triples corrections [non-iterative, ${\cal O}(N_{\bf O}^3 N_{\bf V}^4)$ scaling].
The (T)/[T]  DSRG corrections possess the same computational complexity of the (T) correction in CC theory, however, due to the presence of more terms, they have a slightly higher prefactor.

In the following section, we briefly review the DSRG ansatz and describe various DSRG methods that include connected triples.
Then in Sec.~\ref{sec:results}, we benchmark these DSRG methods on several closed-shell atoms and small molecules.
The results are compared against those obtained by CC and full configuration interaction (FCI).
In Sec.~\ref{sec:conclusion}, we discuss some prospects for generalizing the current formalisms to the MR-DSRG framework.

\section{Theory}
\label{sec:theory}

\subsection{Overview of the SR-DSRG theory}
\label{sec:dsrg}

In this work, we restrict our study of higher excitations to the case of a single Slater determinant reference wave function $\Phi_0$.
The molecular spin orbitals ${\bf G} \equiv \{\phi_p, p = 1, 2, \dots, N_{\bf G}\}$ are classified into sets of occupied ($\bf O$) and virtual ($\bf V$) orbitals of size $N_{\bf O}$ and $N_{\bf V}$, respectively.
The occupied orbitals are labeled by indices $i,j,k,l,\dots$, while virtuals are indicated by $a,b,c,d,\dots$.
We use indices $p,q,r,s,\dots$ to label generic orbitals.
In the SR-DSRG theory, we choose $\Phi_0$ as the Fermi vacuum and all operators are written in normal-ordered form with respect to $\Phi_0$.
For instance, the bare Hamiltonian is expressed as:
\begin{align}
\label{eq:bareH}
\hat{H} = E_0 + \sum_{pq} f_p^q \no{\sqop{p}{q}} + \frac{1}{4} \sum_{pqrs} v_{pq}^{rs} \no{\sqop{pq}{rs}} ,
\end{align}
where $E_0 = \expval{\hat{H}}{\Phi_0}$, $f_p^q = \mel*{\phi_p}{\hat{f}}{\phi_q}$, and $v_{pq}^{rs} = \bra*{\phi_p \phi_q} \!\!\! \ket*{\phi_r \phi_s}$ are the reference energy, Fock matrix elements, and antisymmetrized two-electron integrals, respectively.
A product of second-quantized operators is compactly written as $\sqop{pq\dots}{rs\dots} = \cop{p} \cop{q} \dots \aop{s} \aop{r}$ and curly braces in Eq.~\eqref{eq:bareH} indicate operator normal ordering.

The DSRG transformed Hamiltonian [$\bar{H}(s)$] is given by
\begin{align}
\label{eq:dsrg_trans}
\bar{H}(s) = e^{-\hat{A} (s)} \hat{H} e^{\hat{A} (s)},
\end{align}
where $s$ is the so-called flow parameter, defined in the range $[0, \infty)$.
The operator $\hat{A}(s) = \hat{T}(s) - \hat{T}^\dag (s)$ is defined by an $s$-dependent cluster operator $\hat{T}(s)$.
As in CC theory, $\hat{T}(s)$ is expanded as a sum of $k$-body operators [$\hat{T}_k (s)$]:
\begin{align}
\hat{T}(s) &= \sum_{k=1}^{n} \hat{T}_k (s), \label{eq:T} \\
\hat{T}_k (s) &= \frac{1}{(k!)^2} \sum_{ij\cdots} \sum_{ab\cdots} \tens{t}{ab\cdots}{ij\cdots} (s) \no{\sqop{ab\cdots}{ij\cdots}},
\end{align}
where $n$ can be as large as the total number of electrons.
The DSRG transformed Hamiltonian [$\bar{H}(s)$] is the sum of the correlated DSRG energy, $\bar{H}_0 (s) = \expval{\bar{H} (s)}{\Phi_0}$, and contributions from $k$-body operators [$\bar{H}_k (s)$]:
\begin{align}
\bar{H}(s) &= \bar{H}_0 (s) + \sum_{k=1} \bar{H}_k (s), \label{eq:Hbar} \\
\bar{H}_k (s) &= \frac{1}{(k!)^2} \sum_{pqrs\cdots} \tens{\bar{H}}{rs\cdots}{pq\cdots} (s) \no{\sqop{rs\cdots}{pq\cdots}}.
\end{align}

The DSRG cluster amplitudes are determined by the DSRG flow equation, which consists of a set of many-body conditions:\cite{Datta:2011ca,Evangelista:2014kt}
\begin{align}
\label{eq:flow}
\tens{\bar{H}}{ab\cdots}{ij\cdots} (s) = \tens{r}{ab\cdots}{ij\cdots} (s), \quad ij \in \mathbf{O}, ab \in \mathbf{V}.
\end{align}
The residual $\tens{r}{ab\cdots}{ij\cdots} (s)$ is parameterized in such a way to achieve smooth interpolation between two limits:
i) $\bar{H}(s) = \hat{H}$ when $s=0$ and 
ii) $\tens{\bar{H}}{ab\cdots}{ij\cdots} (s) = 0$ when $s \rightarrow \infty$.
One way to satisfy these requirements is with the following form of $\tens{r}{ab\cdots}{ij\cdots} (s)$:\cite{Evangelista:2014kt}
\begin{align}
\label{eq:source}
\tens{r}{ab\cdots}{ij\cdots} (s) \equiv [ \tens{\bar{H}}{ab\cdots}{ij\cdots} (s) + \Delta^{ij\cdots}_{ab\cdots} \tens{t}{ab\cdots}{ij\cdots} (s) ] e^{-s (\Delta^{ij\cdots}_{ab\cdots})^2},
\end{align}
where $\Delta^{ij\cdots}_{ab\cdots} = \dfock{i} + \dfock{j} + \cdots - \dfock{a} - \dfock{b} - \cdots$ is a M{\o}ller--Plesset denominator defined by the canonical orbital energies $\dfock{p} = f_p^p$.
At this point, the DSRG amplitudes can be solved using Eqs.~\eqref{eq:flow} and \eqref{eq:source} with all instances of $\bar{H} (s)$ replaced by the BCH expansion [Eq.~\eqref{eq:BCH}] written in terms of $\hat{H}$ and $\hat{T} (s)$.

From another perspective, it is easy to see that $\tens{\bar{H}}{ab\cdots}{ij\cdots} (s)$ correspond to the couplings between $\ket*{\Phi_0}$ and the excited determinant $\ket*{\Phi_{ij\cdots}^{ab\cdots}} = \no{\sqop{ab\cdots}{ij\cdots}} \ket*{\Phi_0}$:
\begin{align}
\label{eq:manybody_condition}
\mel*{\Phi_{ij\cdots}^{ab\cdots}}{\bar{H}(s)}{\Phi_0} = \mel*{\Phi_0}{\no{\sqop{ij\cdots}{ab\cdots}} \bar{H}(s)}{\Phi_0} = \tens{\bar{H}}{ab\cdots}{ij\cdots} (s),
\end{align}
where we have used Wick's theorem and the fact that $\braket*{\Phi_{ij\cdots}^{ab\cdots}}{\Phi_0} = 0$.
Therefore, Eqs.~\eqref{eq:flow} and \eqref{eq:source} define a  systematic way to zero the coupling between $\Phi_0$ and its excited configurations.
For a finite value of $s$, only those $\ket*{\Phi_{ij\cdots}^{ab\cdots}}$ with corresponding denominator $|\Delta^{ij\cdots}_{ab\cdots}| > s^{-1/2}$ are decoupled from $\Phi_0$.
In this way, the DSRG ansatz avoids the intruder-state problem caused by small energy denominators.
For brevity, in the following text we will drop the label ``$(s)$'' for all $s$-dependent quantities.

\subsection{Linear truncation schemes: LDSRG(\textit{n})}
\label{sec:truncation}

We now introduce approximations to the DSRG equations and develop a systematically improvable hierarchy containing up to $n$-body terms.
For convenience, we list all the acronyms used in this work in Table \ref{tab:acronyms}.
Following CC theory, we first separate approximate schemes by the level of truncation of the cluster operator [see Eq.~\eqref{eq:T}]. We indicate approximate DSRG schemes containing up to $n$-body substitution operators with the notation DSRG($n$).
For example, the DSRG(2) assumes $\hat{T} \approx \hat{T}_{\{2\}}$ and, for consistency with the amplitude conditions [Eq.~\eqref{eq:source}], the similarity transformed Hamiltonian is approximated as $\bar{H} \approx \bar{H}_{\{2\}}$.
Here, we use the shorthand notation $\hat{O}_{\{n\}} \equiv \sum_{k=0}^{n} \hat{O}_k$ to indicate a generic operator $\hat{O}$ truncated to $n$-body operators.
We also denote truncated cluster operators as $\hat{T}_{\{n\}} \equiv \sum_{k=1}^{n} \hat{T}_k$ and $\hat{A}_{\{n\}} \equiv \hat{T}_{\{n\}} - \hat{T}_{\{n\}}^\dag$.
In the limit of $s \rightarrow \infty$, the DSRG($n$) theory is equivalent to UCC with singles, doubles, $\dots$, up to $n$-tuple excitations.

\begin{table}[!h]
\centering
\ifpreprint
\renewcommand{\arraystretch}{0.7}
\else
\scriptsize
\renewcommand{\arraystretch}{1.25}
\fi
\caption{Summary of the acronyms used in this work.}
\label{tab:acronyms}

\begin{tabular*}{\columnwidth}{@{\extracolsep{\stretch{1}}} *{2}{l} @{}}
\hline
\hline

Acronym & Description \\
\hline

CEPA$_0$ & coupled electron pair approximation variant zero \\

CC & coupled cluster theory \\
\quad CCSD & CC with singles and doubles \\
\quad CCSD(T) & CCSD with perturbative triples \\
\quad CCSDT & CCSD with full triples \\
\quad CCSDT-1 & CCSD with linearized triples  \\
\quad CC3 & approximate CC triples model \\

UCC & unitary coupled cluster theory \\
\quad UCCSD & UCC with singles and doubles \\
\quad UCCSDT & UCCSD with full triples \\

DSRG & driven similarity renormalization group \\

\quad DSRG($n$) & DSRG truncated to $n$-tuple excitations \\
\quad LDSRG($n$) & DSRG($n$) with the $n$-body linear commutator approximation \\
\quad LDSRG(2*) & LDSRG(2) with three-body corrections [Eqs.~\eqref{eq:ildsrg2_2} and \eqref{eq:ildsrg2_3}] \\
\quad qDSRG(2) & LDSRG(2) with recursive quadratic commutators [Eq.~\eqref{eq:q_dsrg2}] \\
\quad LDSRG(3)-$n$ & LDSRG(3) truncated to $(n+3)$-order terms based on $\hat{H}^{(0)}_{\rm Fock}$ \\
\quad LDSRG(3)-$n'$ & LDSRG(3) truncated to $(n+3)$-order terms based on $\hat{H}^{(0)}_{\rm Fink}$ \\
\quad LDSRG(3;C$k$) & LDSRG(3) with 3-body terms truncated to $k$-nested commutators \\
\quad [T] & perturbative triples defined by Eq.~\eqref{eq:dsrg[t]} \\
\quad (T) & perturbative triples defined by Eq.~\eqref{eq:dsrg(t)} \\

\hline
\hline
\end{tabular*}
\end{table}

Second, we classify truncated schemes according to the approximation of the nested commutators in the BCH series.
In the linear $i$-body approximation [Eq.~\eqref{eq:rsc}], every single commutator contains at most $i$-body term.
Thus, the transformed Hamiltonian that includes $(k + 1)$-nested commutator [$\hat{O}^{k+1}_{\{i\}}$] can be computed recursively:
\begin{align}
\label{eq:BCH_recursive}
\hat{O}^{k+1}_{\{i\}} = \hat{O}^{k}_{\{i\}} +  \frac{1}{k + 1}[\hat{O}^{k}_{\{i\}} - \hat{O}^{k-1}_{\{i\}}, \hat{A}]_{\{i\}}, \quad k = 1, 2, 3, \dots
\end{align}
starting from $\hat{O}^{0} = \hat{H}$ and $\hat{O}^{1}_{\{i\}} = \hat{H} + [\hat{H}, \hat{A}]_{\{i\}}$.
This many-body truncation scheme can be extended beyond the linear commutator level.
For instance, the quadratic $i$-body approximation of Neuscamman \textit{et al.} assumes $[[\,\cdot\, , \hat{A}],\hat{A}] \approx [[\,\cdot\, , \hat{A}],\hat{A}]_{\{i\}}$,\cite{Neuscamman:2009cy} where $[\,\cdot\, , \hat{A}]$ is computed exactly.

\begin{table}[!h]
\centering
\ifpreprint
\renewcommand{\arraystretch}{0.75}
\else
\renewcommand{\arraystretch}{1.25}
\fi
\caption{Terms included in the L2SD truncation scheme that arise from the commutator $[\hat{O}_{\{2\}}, \hat{T}_{\{2\}}]_{\{2\}}$. Einstein convention of summation over repeated indices is assumed. The index permutation operator is indicated by $\permop(p/q) \no{\qop{p} \qop{q}} = \no{\qop{p} \qop{q}} - \no{\qop{q} \qop{p}}$, where $\qop{}$ is a generic second-quantized operator ($\cop{}$ or $\aop{}$).}
\label{tab:l2sd}

\begin{tabular*}{\columnwidth}{@{\extracolsep{\stretch{1}}} *{3}{l} @{}}
\hline
\hline

Term & Expression & Cost \\
\hline

0-1 & $+ \tens{O}{i}{a} \tens{t}{a}{i}$ & $N_{\bf O} N_{\bf V}$ \\
0-2 & $+ \frac{1}{4} \tens{O}{ij}{ab} \tens{t}{ab}{ij} $ & $N_{\bf O}^2 N_{\bf V}^2$  \\

\hline

1-1a & $+ \tens{O}{p}{a} \tens{t}{a}{i} \nsqop{p}{i} $ & $N_{\bf O} N_{\bf V} N_{\bf G}$ \\
1-1b & $- \tens{O}{i}{p} \tens{t}{a}{i} \nsqop{a}{p}$ & $N_{\bf O} N_{\bf V} N_{\bf G}$ \\
1-2 & $+ \tens{O}{j}{b} \tens{t}{ab}{ij} \nsqop{a}{i}$ & $N_{\bf O}^2 N_{\bf V}^2$ \\
1-3 & $+ \tens{O}{pi}{qa} \tens{t}{a}{i} \nsqop{p}{q}$ & $N_{\bf O} N_{\bf V} N_{\bf G}^2$ \\
1-4a & $+ \frac{1}{2} \tens{O}{rj}{ab} \tens{t}{ab}{ij} \nsqop{r}{i}$ & $N_{\bf O}^2 N_{\bf V}^2 N_{\bf G}$ \\
1-4b & $- \frac{1}{2} \tens{O}{ij}{pb} \tens{t}{ab}{ij} \nsqop{a}{p}$ & $N_{\bf O}^2 N_{\bf V}^2 N_{\bf G}$ \\

\hline

2-1a & $+ \frac{1}{4} {\cal P} (p/b) \tens{O}{p}{a} \tens{t}{ab}{ij} \nsqop{pb}{ij}$ & $N_{\bf O}^2 N_{\bf V}^2 N_{\bf G}$ \\
2-1b & $- \frac{1}{4} {\cal P} (q/j) \tens{O}{i}{q} \tens{t}{ab}{ij} \nsqop{ab}{qj}$ & $N_{\bf O}^2 N_{\bf V}^2 N_{\bf G}$ \\
2-2a & $+ \frac{1}{4} {\cal P} (r/i) \tens{O}{pq}{ar} \tens{t}{a}{i} \nsqop{pq}{ir}$ & $N_{\bf O} N_{\bf V} N_{\bf G}^3$ \\
2-2b & $- \frac{1}{4} {\cal P} (p/a) \tens{O}{pi}{rs} \tens{t}{a}{i} \nsqop{pa}{rs}$ & $N_{\bf O} N_{\bf V} N_{\bf G}^3$ \\
2-3a & $+ \frac{1}{8} \tens{O}{pq}{ab} \tens{t}{ab}{ij} \nsqop{pq}{ij}$ & $N_{\bf O}^2 N_{\bf V}^2 N_{\bf G}^2$ \\
2-3b & $+ \frac{1}{8} \tens{O}{ij}{pq} \tens{t}{ab}{ij} \nsqop{ab}{pq}$ & $N_{\bf O}^2 N_{\bf V}^2 N_{\bf G}^2$ \\
2-3c & $+ \frac{1}{4} {\cal P}(p/b) {\cal P}(q/j) \tens{O}{ip}{aq} \tens{t}{ab}{ij} \nsqop{pb}{qj}$ & $N_{\bf O}^2 N_{\bf V}^2 N_{\bf G}^2$ \\

\hline
\hline
\end{tabular*}
\end{table}

The LDSRG(2) approach has been introduced in Ref.~\citenum{Evangelista:2014kt} and extended to the multireference formalism.\cite{Li:2016hb}
The LDSRG(2) equations are very simple and they are reproduced in Table \ref{tab:l2sd}.
Note that we only need to derive the expressions for $[\hat{O}_{\{2\}}, \hat{T}]_{\{2\}}$ because of the recursive algorithm to evaluate $\bar{H}$ [Eq.~\eqref{eq:BCH_recursive}] and the fact that $[\hat{O}^{k}_{\{2\}}, \hat{A}] = [\hat{O}^{k}_{\{2\}}, \hat{T}] + [\hat{O}^{k}_{\{2\}}, \hat{T}]^\dag$.
In the LDSRG(2), the most expensive terms are those from the commutator $[\hat{O}_2, \hat{T}_2]_2$ (terms 2-3 in Table \ref{tab:l2sd}), which scale as ${\cal O}(N_{\bf O}^2 N_{\bf V}^2 N_{\bf G}^2)$.
Although this cost is similar to that of CCSD [${\cal O}(N_{\bf O}^2 N_{\bf V}^4)$], this term must be evaluated for each step in the recursive computation of $\bar{H}$.

\begin{table}[!h]
\begin{threeparttable}
\centering
\ifpreprint
\renewcommand{\arraystretch}{0.7}
\else
\renewcommand{\arraystretch}{1.25}
\fi

\caption{Terms included in the L3SDT scheme that arise from the commutator $[\hat{O}_{\{3\}}, \hat{T}_{\{3\}}]_{\{3\}}$. The terms included in the L2SD are not shown. Einstein convention of summation over repeated indices is assumed. The index permutation operator is indicated by $\permop (p/q) \no{\qop{p} \qop{q}} = \no{\qop{p} \qop{q}} - \no{\qop{q} \qop{p}}$ and $\permop (p/rs) \no{\qop{p} \qop{r} \qop{s}} = \no{\qop{p} \qop{r} \qop{s}} - \no{\qop{r} \qop{p} \qop{s}} - \no{\qop{s} \qop{r} \qop{p}}$, where $\qop{}$ is a generic second-quantized operator ($\cop{}$ or $\aop{}$).}
\label{tab:l3sdt}

\begin{tabular*}{\columnwidth}{@{\extracolsep{\stretch{1}}} l l l @{}}
\hline
\hline

Term & Expression & Cost \\
\hline
0-3 & $+ \frac{1}{36} \tens{O}{ijk}{abc} \tens{t}{abc}{ijk} $ & $N_{\bf O}^3 N_{\bf V}^3$ \\

\hline

1-5 & $
+\frac{1}{4}  \tens{O}{j k}{b c} \tens{t}{a b c}{i j k} \nsqop{a}{i} $ & $N_{\bf O}^3 N_{\bf V}^3$ \\
1-6 & $
+\frac{1}{4}  \tens{O}{p i j}{q a b} \tens{t}{a b}{i j} \nsqop{p}{q} $ & $N_{\bf O}^2 N_{\bf V}^2 N_{\bf G}^2$ \\
1-7a & $
+\frac{1}{12}  \tens{O}{p j k}{a b c} \tens{t}{a b c}{i j k} \nsqop{p}{i} $ & $N_{\bf O}^3 N_{\bf V}^3 N_{\bf G}$ \\
1-7b & $
-\frac{1}{12}  \tens{O}{i j k}{p b c} \tens{t}{a b c}{i j k} \nsqop{a}{p} $ & $N_{\bf O}^3 N_{\bf V}^3 N_{\bf G}$ \\

\hline

2-4 & $
+\frac{1}{4}  \tens{O}{k}{c} \tens{t}{a b c}{i j k} \nsqop{a b}{i j} $ & $N_{\bf O}^3 N_{\bf V}^3$ \\
2-5a & $
+\frac{1}{8} {\cal P}(p / i) \tens{O}{j k}{p c} \tens{t}{a b c}{i j k} \nsqop{a b}{p i} $ & $N_{\bf O}^3 N_{\bf V}^3 N_{\bf G}$ \\
2-5b & $
-\frac{1}{8} {\cal P}(p / a) \tens{O}{p k}{b c} \tens{t}{a b c}{i j k} \nsqop{p a}{i j} $ & $N_{\bf O}^3 N_{\bf V}^3 N_{\bf G}$ \\
2-6 & $
+\frac{1}{4}  \tens{O}{p q i}{r s a} \tens{t}{a}{i} \nsqop{p q}{r s} $ & $N_{\bf O} N_{\bf V} N_{\bf G}^4$ \\
2-7a & $
+\frac{1}{8} {\cal P}(r / i) \tens{O}{p q j}{r a b} \tens{t}{a b}{i j} \nsqop{p q}{r i} $ & $N_{\bf O}^2 N_{\bf V}^2 N_{\bf G}^3$ \\
2-7b & $
-\frac{1}{8} {\cal P}(p / a) \tens{O}{p i j}{q r b} \tens{t}{a b}{i j} \nsqop{p a}{q r} $ & $N_{\bf O}^2 N_{\bf V}^2 N_{\bf G}^3$ \\
2-8a & $
+\frac{1}{24}  \tens{O}{p q k}{a b c} \tens{t}{a b c}{i j k} \nsqop{p q}{i j} $ & $N_{\bf O}^3 N_{\bf V}^3 N_{\bf G}^2$ \\
2-8b & $
+\frac{1}{24}  \tens{O}{i j k}{p q c} \tens{t}{a b c}{i j k} \nsqop{a b}{p q} $ & $N_{\bf O}^3 N_{\bf V}^3 N_{\bf G}^2$ \\
2-8c & $
+\frac{1}{16} {\cal P}(p / a){\cal P}(q / i) \tens{O}{p j k}{q b c} \tens{t}{a b c}{i j k} \nsqop{p a}{q i} $ & $N_{\bf O}^3 N_{\bf V}^3 N_{\bf G}^2$ \\

\hline

3-1a & $
+\frac{1}{36} {\cal P}(p / a b) \tens{O}{p}{c} \tens{t}{a b c}{i j k} \nsqop{p a b}{i j k} $ & $N_{\bf O}^3 N_{\bf V}^3 N_{\bf G}$ \\
3-1b & $
-\frac{1}{36} {\cal P}(p / i j) \tens{O}{k}{p} \tens{t}{a b c}{i j k} \nsqop{a b c}{p i j} $ & $N_{\bf O}^3 N_{\bf V}^3 N_{\bf G}$ \\
3-2a & $
-\frac{1}{36} {\cal P}(p q / a){\cal P}(r / i j) \tens{O}{p q}{r b} \tens{t}{a b}{i j} \nsqop{p q a}{r i j} $ & $N_{\bf O}^2 N_{\bf V}^2 N_{\bf G}^3$ \\
3-2b & $
+\frac{1}{36} {\cal P}(p / a b){\cal P}(q r / i) \tens{O}{p j}{q r} \tens{t}{a b}{i j} \nsqop{p a b}{q r i} $ & $N_{\bf O}^2 N_{\bf V}^2 N_{\bf G}^3$ \\
3-3a & $
+\frac{1}{72} {\cal P}(p q / i) \tens{O}{j k}{p q} \tens{t}{a b c}{i j k} \nsqop{a b c}{p q i} $ & $N_{\bf O}^3 N_{\bf V}^3 N_{\bf G}^2$ \\
3-3b & $
+\frac{1}{72} {\cal P}(p q / a) \tens{O}{p q}{b c} \tens{t}{a b c}{i j k} \nsqop{p q a}{i j k} $ & $N_{\bf O}^3 N_{\bf V}^3 N_{\bf G}^2$ \\
3-3c & $
+\frac{1}{36} {\cal P}(p / a b){\cal P}(q / i j) \tens{O}{p k}{q c} \tens{t}{a b c}{i j k} \nsqop{p a b}{q i j} $ & $N_{\bf O}^3 N_{\bf V}^3 N_{\bf G}^2$ \\
3-4a & $
+\frac{1}{36} {\cal P}(s t / i) \tens{O}{p q r}{s t a} \tens{t}{a}{i} \nsqop{p q r}{s t i} $ & $N_{\bf O} N_{\bf V} N_{\bf G}^5$ \\
3-4b & $
-\frac{1}{36} {\cal P}(p q / a) \tens{O}{p q i}{r s t} \tens{t}{a}{i} \nsqop{p q a}{r s t} $ & $N_{\bf O} N_{\bf V} N_{\bf G}^5$ \\
3-5a & $
+\frac{1}{72} {\cal P}(s / i j) \tens{O}{p q r}{s a b} \tens{t}{a b}{i j} \nsqop{p q r}{s i j} $ & $N_{\bf O}^2 N_{\bf V}^2 N_{\bf G}^4$ \\
3-5b & $
+\frac{1}{72} {\cal P}(p / a b) \tens{O}{p i j}{q r s} \tens{t}{a b}{i j} \nsqop{p a b}{q r s} $ & $N_{\bf O}^2 N_{\bf V}^2 N_{\bf G}^4$ \\
3-5c & $
+\frac{1}{36} {\cal P}(p q / a){\cal P}(r s / i) \tens{O}{p q j}{r s b} \tens{t}{a b}{i j} \nsqop{p q a}{r s i} $ & $N_{\bf O}^2 N_{\bf V}^2 N_{\bf G}^4$ \\
3-6a & $
+\frac{1}{216}  \tens{O}{p q r}{a b c} \tens{t}{a b c}{i j k} \nsqop{p q r}{i j k} $ & $N_{\bf O}^3 N_{\bf V}^3 N_{\bf G}^3$ \\
3-6b & $
-\frac{1}{216}  \tens{O}{i j k}{p q r} \tens{t}{a b c}{i j k} \nsqop{a b c}{p q r} $ & $N_{\bf O}^3 N_{\bf V}^3 N_{\bf G}^3$ \\
3-6c & $
-\frac{1}{72} {\cal P}(p q / a){\cal P}(r / i j) \tens{O}{p q k}{r b c} \tens{t}{a b c}{i j k} \nsqop{p q a}{r i j} $ & $N_{\bf O}^3 N_{\bf V}^3 N_{\bf G}^3$ \\
3-6d & $
+\frac{1}{72} {\cal P}(p / a b){\cal P}(q r / i) \tens{O}{p j k}{q r c} \tens{t}{a b c}{i j k} \nsqop{p a b}{q r i} $ & $N_{\bf O}^3 N_{\bf V}^3 N_{\bf G}^3$ \\

\hline
\hline
\end{tabular*}
\end{threeparttable}
\end{table}

\begin{figure}[!h]
\ifpreprint
    \includegraphics[width=0.50\columnwidth]{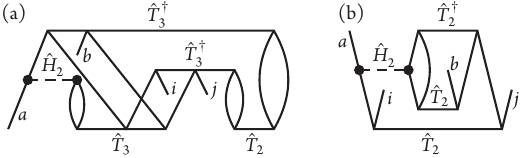}
\else
    \includegraphics[width=0.9\columnwidth]{ucc.pdf}
\fi
\caption{Antisymmetrized Goldstone skeleton diagrams for (a) one of the ${\cal O}(N_{\bf O}^3 N_{\bf V}^6)$ terms found in UCCSDT and (b) one of the ${\cal O}(N_{\bf V}^6)$ terms found in UCCSD. Open lines are labeled with occupied or virtual indices.}
\label{fig:uccsdt}
\end{figure}

Continuing on this route, the simplest way to introduce triple excitations is via the linearized truncation scheme (L3SDT), defined by $\hat{T} \approx \hat{T}_{\{3\}}$ and the commutator approximation $[\,\cdot\, , \hat{A}] \approx [\,\cdot\, , \hat{A}]_{\{3\}}$.
As shown in Table~\ref{tab:l3sdt}, the resulting LDSGR(3) equations include a number of additional terms compared to the LDSRG(2).
An inspection of these expressions reveals that computing $[\hat{O}_3, \hat{T}_3]_3$ (terms 3-6 in Table \ref{tab:l3sdt}) scales as ${\cal O}(N_{\bf O}^3 N_{\bf V}^3 N_{\bf G}^3)$, a cost that is significantly higher than that of CCSDT [${\cal O}(N_{\bf O}^3 N_{\bf V}^5)$].
In fact, the asymptotic scaling of LDSRG($n$) is ${\cal O}(N_{\bf O}^n N_{\bf V}^n N_{\bf G}^n)$, which in comparison to CC with $n$-tuple excitations becomes significantly more costly when $n > 2$.
Note that the high computational cost of LDSRG($n$) is a direct consequence of using a unitary ansatz [Eq.~\eqref{eq:BCH}], which also affects UCC theories.
For example, Fig.~\ref{fig:uccsdt}(a) shows an antisymmetrized Goldstone diagram that arises from the 4-nested commutator and contributes to both LDSRG(3) and unitary CCSDT (UCCSDT) theories.
This term scales as ${\cal O}(N_{\bf O}^3 N_{\bf V}^6)$ and requires forming an intermediate of size $N_{\bf O} N_{\bf V}^5$.

\subsection{Iterative approximations to the LDSRG(3)}
\label{sec:ldsrg3-n}

\subsubsection{Approximate triples methods: LDSRG(3)-n and LDSRG(3)-n$'$ (n = 1, 2, 3, 4)}

To obtain a more affordable iterative triples method, we now consider methods which include a subset of the LDSRG(3) terms reported in Table \ref{tab:l3sdt}.
In order to decide which terms to retain, we use perturbation theory to assign an order to each contribution.
Specifically, we consider two types of zeroth-order Hamiltonians.
The first choice contains the diagonal blocks of the Fock operator:
\begin{align}
\label{eq:H0th_Fock}
\hat{H}^{(0)}_{\rm Fock} = E_0 + \sum_{ij} f_i^j \no{\sqop{i}{j}} + \sum_{ab} f_a^b \no{\sqop{a}{b}},
\end{align}
and the first-order Hamiltonian is then given by $\hat{H}^{(1)}_{\rm Fock} = \hat{H} - \hat{H}^{(0)}_{\rm Fock}$.
If we do not assume a HF reference wave function, both $\hat{T}_1$ and $\hat{T}_2$ are first-order quantities, while $\hat{T}_3$ enters at second order in the perturbation.
We also recognize that the three-body intermediates in LDSRG(3) first appear at second order, resulting from $[\hat{H}^{(1)}_{\rm Fock}, \hat{A}_2]_3$.
An order-by-order perturbative analysis shows that the L3SDT approximation introduces errors to the fifth-order energy due to the lack of induced four-body interactions.
Overall, the LDSRG(3) energy is complete through fourth order and the wave function is exact through second order.

Another zeroth-order Hamiltonian considered here is Fink's retaining-excitation Hamiltonian:\cite{Fink:2006gu}
\begin{align}
\label{eq:H0th_Fink}
\hat{H}^{(0)}_{\rm Fink} =&\, \hat{H}^{(0)}_{\rm Fock} + \frac{1}{4} \sum_{ijkl} \tens{v}{ij}{kl} \nsqop{ij}{kl} + \frac{1}{4} \sum_{abcd} \tens{v}{ab}{cd} \nsqop{ab}{cd} \notag \\
&+ \frac{1}{4} \sum_{ij}\sum_{ab} \tens{v}{ia}{jb} \left( \nsqop{ia}{jb} - \nsqop{ai}{jb} - \nsqop{ia}{bj} + \nsqop{ai}{bj} \right).
\end{align}
When applied to a determinant with $k$ excited electrons (with respect to the reference), $\hat{H}^{(0)}_{\rm Fink}$ does not change its excitation level.
It can be shown that the perturbation orders of $\hat{T}_1$, $\hat{T}_2$, and $\hat{T}_3$ using $\hat{H}^{(0)}_{\rm Fink}$ are identical to those obtained with $\hat{H}^{(0)}_{\rm Fock}$.
However, the three-body intermediates associated with the operators $\nsqop{ijk}{abl}$, $\nsqop{abl}{ijk}$, $\nsqop{ijd}{abc}$, $\nsqop{abc}{ijd}$ become first-order quantities due to contractions of $[\hat{H}^{(0)}_{\rm Fink}, \hat{A}_2]_3$.
Consequently, quadruple excitations $\hat{T}_4$ appear at the second order of perturbation, ignoring which yields errors to the fourth-order energy.
Nonetheless, Fink's Hamiltonian contains an important subset of two-electron integrals (e.g., $\tens{v}{ij}{ij}$ and $\tens{v}{ia}{ia}$) that are larger in magnitude than the $\tens{v}{ij}{ab}$ type integrals.

We now classify the LDSRG(3) terms (Table \ref{tab:l3sdt}) according to their lowest order contribution to the energy.
This order of perturbation is calculated as the sum of three components:
i) the lowest order of the intermediate $\hat{O}$,
ii) the order of cluster operator $\hat{T}$,
and iii) the number of additional operators $\hat{A}_k\, (k = 1, 2)$ needed to close the corresponding diagram of $[\hat{O}, \hat{T}]$.
For example, term 2-4, $\frac{1}{4} \tens{O}{k}{c} \tens{t}{a b c}{i j k} \nsqop{a b}{i j}$, contributes to the fourth-order energy because $\tens{O}{k}{c}$ can be a first-order quantity and $\tens{t}{abc}{ijk}$ is of second-order, and one $\hat{T}_2^\dag$ is required to be fully contracted.
As such, the LDSRG(3) terms yield fourth- through eighth-order energies, suggesting a sequence of levels of theories.
We thus introduce the LDSRG(3)-$n$ ($n=1,2,3,4$) methods by including those terms in Table \ref{tab:l3sdt} that contribute to the ($n+3$)-order energy in perturbation theory based on $\hat{H}^{(0)}_{\rm Fock}$ [Eq.~\eqref{eq:H0th_Fock}].
The LDSRG(3)-$n'$ theories are defined in a similar way except that $\hat{H}^{(0)}_{\rm Fink}$ [Eq.~\eqref{eq:H0th_Fink}] is used instead.

\begin{figure}[!ht]
\centering
\ifpreprint
    \includegraphics[width=0.50\columnwidth]{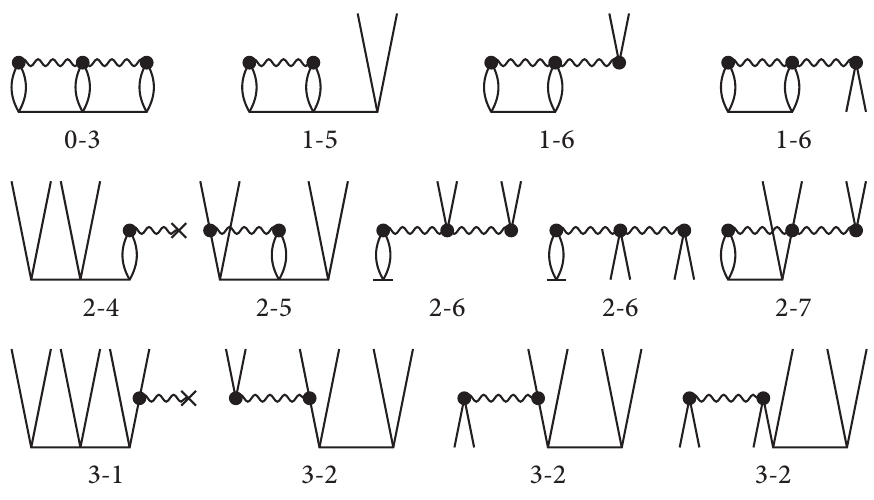}
\else
    \includegraphics[width=0.90\columnwidth]{L3SDT_PT4}
\fi
\caption{Antisymmetrized Goldstone skeleton diagrams of LDSRG(3) that affect the fourth-order energy based on $\hat{H}^{(0)}_{\rm Fock}$ or $\hat{H}^{(0)}_{\rm Fink}$. Here, we use wiggly lines to indicate the intermediate $\hat{O}$ for each level of nested commutator in the BCH expansion, while $\hat{T}$ is indicated by horizontal solid lines. Note that multiple diagrams may contribute to the same algebraic term labeled in Table \ref{tab:l3sdt}. For example, the three 3-2 diagrams come from terms 3-2a and 3-2b in Table \ref{tab:l3sdt}.}
\label{fig:ldsrg3_pt4}
\end{figure}

\begin{figure}[!ht]
\centering
\ifpreprint
    \includegraphics[width=0.50\columnwidth]{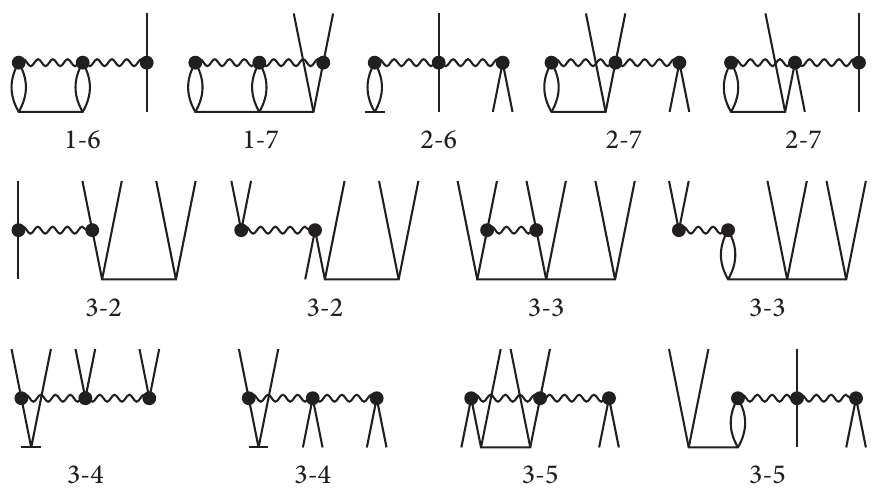}
\else
    \includegraphics[width=0.90\columnwidth]{L3SDT_PT4_5}
\fi
\caption{Antisymmetrized Goldstone skeleton diagrams of LDSRG(3) that affect the fourth-order energy based on $\hat{H}^{(0)}_{\rm Fink}$ or fifth-order energy based on $\hat{H}^{(0)}_{\rm Fock}$. See the caption of Fig.~\ref{fig:ldsrg3_pt4} for details.}
\label{fig:ldsrg3_pt4.5}
\end{figure}

Figure \ref{fig:ldsrg3_pt4} shows the diagrams considered in the LDSRG(3)-1 ansatz that contribute to the commutator $[\hat{O},\hat{T}]$.
An inspection of these diagrams suggests that the overall scaling of LDSRG(3)-1 is ${\cal O} (N_{\bf O}^3 N_{\bf V}^4)$, resulting from terms 2-5, 2-6, 2-7, and 3-2.
Note that this cost is identical to that of the CCSDT-1\cite{Lee:1984fq,*Lee:1998gm,Urban:1985bq} and CC3\cite{Koch:1997eu} methods.
The LDSRG(3)-1 scheme requires storing a three-body intermediate of size $N_{\bf O}^2 N_{\bf V}^4$ (see terms 2-7 and 3-2), a significant advantage over the LDSRG(3) theory ($N_{\bf G}^6$ storage cost).
In Fig.~\ref{fig:ldsrg3_pt4.5}, we present the additional diagrams included in the LDSRG(3)-1$'$ scheme.
In comparison to LDSRG(3)-1, the computational cost is now dominated by term 3-5 [${\cal O} (N_{\bf O}^3 N_{\bf V}^5)$], while the storage cost remains the same.

\begin{figure}[!ht]
\centering
\ifpreprint
    \includegraphics[width=0.50\columnwidth]{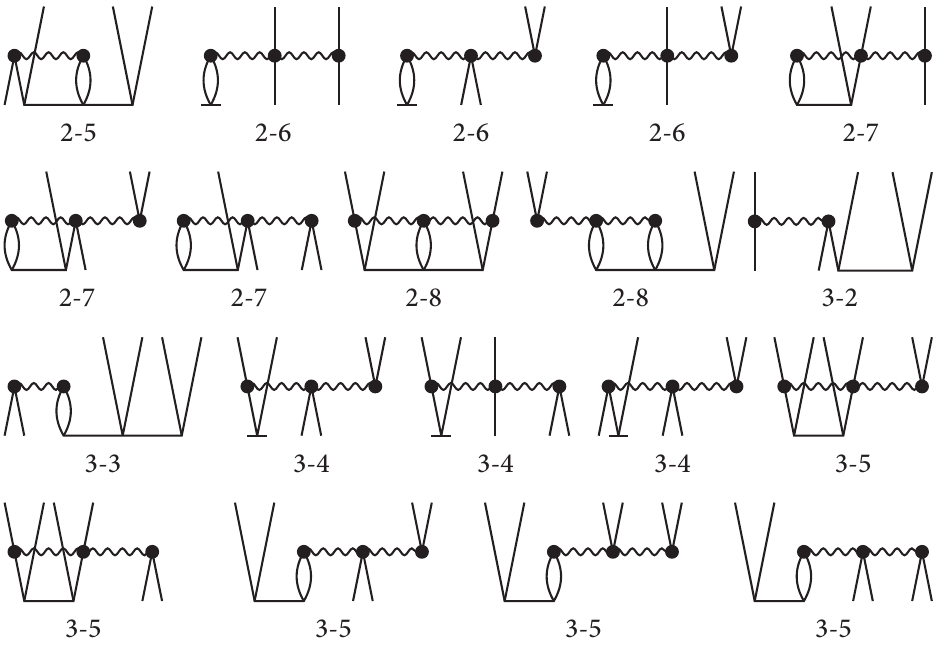}
\else
    \includegraphics[width=0.90\columnwidth]{L3SDT_PT5}
\fi
\caption{Antisymmetrized Goldstone skeleton diagrams of LDSRG(3) that affect the fifth-order energy based on $\hat{H}^{(0)}_{\rm Fock}$ or $\hat{H}^{(0)}_{\rm Fink}$. See the caption of Fig.~\ref{fig:ldsrg3_pt4} for details.}
\label{fig:ldsrg3_pt5}
\end{figure}

\begin{figure}[!ht]
\centering
\ifpreprint
    \includegraphics[width=0.50\columnwidth]{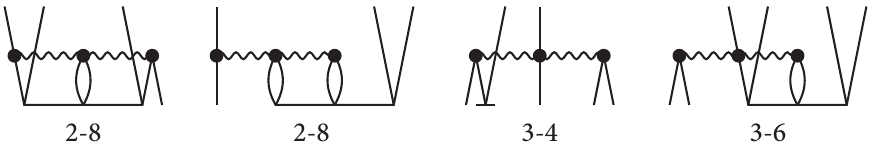}
\else
    \includegraphics[width=0.90\columnwidth]{L3SDT_PT5_6}
\fi
\caption{Antisymmetrized Goldstone skeleton diagrams of LDSRG(3) that affect the fifth-order energy based on $\hat{H}^{(0)}_{\rm Fink}$ or sixth-order energy based on $\hat{H}^{(0)}_{\rm Fock}$. See the caption of Fig.~\ref{fig:ldsrg3_pt4} for details.}
\label{fig:ldsrg3_pt5.5}
\end{figure}

We can continue this route and obtain the LDSRG(3)-2 (and -2$'$) theory by including fifth-order terms.
The resulting diagrams are plotted in Figs.~\ref{fig:ldsrg3_pt5} and \ref{fig:ldsrg3_pt5.5}.
Note that the three-body intermediate is now of size $N_{\bf O} N_{\bf V}^5$ for both approaches (see terms 2-6, 2-7, 2-8, 3-4, 3-5 of Fig.~\ref{fig:ldsrg3_pt5}).
The computational cost for LDSRG(3)-2 scales as ${\cal O} (N_{\bf O}^3 N_{\bf V}^5)$, while for LDSRG(3)-2$'$ it is ${\cal O} (N_{\bf O}^4 N_{\bf V}^5)$.

Including sixth and higher-order terms leads to formulations that require identical computational resources as the complete LDSRG(3) theory.
As such, no practical benefits are gained to employ the LDSRG(3)-3 and -4 methods.
For completeness, we report these higher-order diagrams in the Supplementary Material.
The computational and storage cost of all DSRG methods are summarized in Table \ref{tab:cost}.
Note that all the variants of LDSRG(3) have a storage cost that is equal or higher than that of CCSDT [${\cal O} (N_{\bf O}^3 N_{\bf V}^3)$].

\begin{table}[ht!]
\begin{threeparttable}
\centering
\ifpreprint
\renewcommand{\arraystretch}{1.0}
\else
\renewcommand{\arraystretch}{1.5}
\fi
\caption{Computational and storage costs (in big $\cal O$ notation) of all LDSRG(3) variants considered in this work. }
\label{tab:cost}
\begin{tabular*}{\columnwidth}{@{\extracolsep{\fill} } l c c c c @{}}
\hline
\hline
 & \multicolumn{2}{c}{Original} & \multicolumn{2}{c}{Truncated $\bar{H}_3$ [Eq.~\eqref{eq:H3ncomm2}]} \\
\cline{2-3} \cline{4-5}
Method & \multicolumn{1}{c}{comput.} & \multicolumn{1}{c}{storage} & \multicolumn{1}{c}{comput.} & \multicolumn{1}{c}{storage} \\
\hline
LDSRG(3)-1 & $N_{\bf O}^3 N_{\bf V}^4$ & $N_{\bf O}^2 N_{\bf V}^4$ & $N_{\bf O}^3 N_{\bf V}^4$ & $N_{\bf O}^3 N_{\bf V}^3$ \\
LDSRG(3)-1$'$ & $N_{\bf O}^3 N_{\bf V}^5$ & $N_{\bf O}^2 N_{\bf V}^4$ & $N_{\bf O}^4 N_{\bf V}^4$ & $N_{\bf O}^3 N_{\bf V}^3$ \\
LDSRG(3)-2 & $N_{\bf O}^3 N_{\bf V}^5$ & $N_{\bf O} N_{\bf V}^5$ & $N_{\bf O}^3 N_{\bf V}^5$ & $N_{\bf O}^3 N_{\bf V}^3$ \\
LDSRG(3)-2$'$ & $N_{\bf O}^4 N_{\bf V}^5$ & $N_{\bf O} N_{\bf V}^5$ & $N_{\bf O}^3 N_{\bf V}^5$ & $N_{\bf O}^3 N_{\bf V}^3$ \\
LDSRG(3) & $N_{\bf O}^3 N_{\bf V}^6$ & $N_{\bf V}^6$ & $N_{\bf O}^4 N_{\bf V}^5$ & $N_{\bf O}^2 N_{\bf V}^4$ \\
\hline
\hline
\end{tabular*}
\end{threeparttable}
\end{table}

\subsubsection{Three-body intermediates truncation}

In this section, we investigate how truncating the BCH expansion affects the energy in the LDSRG(3) framework.
This approximation is motivated by observing that the ${\cal O}(N_{\bf O}^3 N_{\bf V}^6)$ terms of LDSRG(3) [e.g., Fig.~\ref{fig:uccsdt}(a)] first appear at the 4-nested commutator.
If the BCH series can be terminated early without losing much accuracy, we can not only derive a closed form of the energy and amplitudes, but potentially lower the computational scaling and storage costs (see Table \ref{tab:cost}).

Previous work on unitary CCSD (UCCSD) has shown that four nested commutators are necessary to reach sub-m\Eh accuracy.\cite{Evangelista:2011jp}
Here, we expect a faster convergence to chemical accuracy due to smaller energy contributions from triples.
To this end, we approximate $\bar{H}_3$ by terminating the BCH series at $m$-nested commutators ($m = 1, 2, 3, 4$).
For example, when $m = 2$, $\bar{H}_3$ is given by
\begin{align}
\label{eq:H3ncomm2}
\bar{H}_3 \approx [\hat{H}, \hat{A}_{\{3\}}]_3 + \frac{1}{2} [[\hat{H}, \hat{A}_{\{3\}}]_{\{3\}}, \hat{A}_{\{3\}}]_3.
\end{align}
In the following, we denote the LDSRG(3) with $\bar{H}_3$ truncated at the $m$-nested commutator as LDSRG(3;C$m$).
Note that $\bar{H}_1$ and $\bar{H}_2$ are not explicitly truncated here, but they are implicitly affected due to the recursive L3SDT algorithm.

\begin{table}[h!]
\begin{threeparttable}
\centering
\ifpreprint
\renewcommand{\arraystretch}{1.0}
\else
\renewcommand{\arraystretch}{1.2}
\fi
\caption{Statistics of energy deviations (in m\Eh) from the LDSRG(3)-1/6-31G values for terminating its BCH expansion at the $m$-nested commutator ($m = 1, 2, 3, 4$) for $\bar{H}_3$.
Statistics were calculated among twenty-eight small closed-shell atoms and molecules (see Sec.~\ref{sec:results}).
The flow parameter is set to $s = 10^3$ \sunit.
}
\label{table:comm}
\begin{tabular*}{\columnwidth}{@{\extracolsep{\stretch{1}}} l *{2}{d{1.3}} *{2}{d{2.3}} @{}}
\hline
\hline
Statistics & \multicolumn{1}{c}{1} & \multicolumn{1}{c}{2} & \multicolumn{1}{c}{3} & \multicolumn{1}{c}{4} \\
\hline
Mean signed error & 1.922 & 0.279 & -0.039 & -0.006 \\
Mean absolute error & 1.922 & 0.279 & 0.039 & 0.006 \\
Standard deviation & 2.106 & 0.855 & 0.144 & 0.029 \\
\hline
\hline
\end{tabular*}
\end{threeparttable}
\end{table}

Table \ref{table:comm} shows the error of terminating the BCH series for $\bar{H}_3$ in the LDSRG(3)-1 scheme.
In general, the error reduces an order of magnitude when an extra nested commutator is considered.
A large 1.92 m\Eh mean absolute error (MAE) is observed when we keep only the linear term [e.g., $\bar{H}_3 \approx [\hat{H}, \hat{A}_2]_3$ for LDSRG(3)-1].
The truncation error becomes essentially negligible if the BCH formula is terminated at the four-nested commutator term.
Since we aim to achieve sub-chemical accuracy with the LDSRG(3) theory, it is already sufficient to keep only the linear and quadratic terms of $\bar{H}_3$, as shown in Eq.~\eqref{eq:H3ncomm2}.
Other LDSRG(3;C2) variants also yield sub-m\Eh error [e.g., MAE = 0.21 m\Eh for LDSRG(3;C2)-2] with respect to the corresponding untruncated approaches.

\subsection{Perturbative approximations to the LDSRG(3)}
\label{subsec:pt}

\subsubsection{qDSRG(2) and LDSRG(2*)}

One of the major goals in this work is to propose a practical perturbative triples correction to the unitary DSRG(2) formalism.
However, the strategy of adding (T) corrections on top of CCSD cannot be directly applied to the LDSRG(2) approach because the latter neglects some important fourth-order energy terms.
Numerical evidence (see Sec.~\ref{sec:results}) shows that these missing terms cause the LDSRG(2) method to overestimate the correlation energy and yield energies that are closer to FCI, similar to the case of the coupled electron pair approximation (CEPA).\cite{Meyer:1973fb,Koch:1981dx}

To address this issue, we consider approximate methods that account for the missing fourth-order terms in the LDSRG(2) Hamiltonian.
The strategy we follow in this work is to take the LDSRG(3)-$n$ and LDSRG(3)-$n'$ methods  in their lowest order approximation ($n = 1$) and ignore the contributions due to the three-body amplitudes.
The scheme that offers the best compromise between accuracy and cost is based on the LDSRG(3)-1 (with $\hat{T}_3 = 0$), which we refer to as pseudo-quadratic DSRG(2) [qDSRG(2)].
This method includes only terms 1-6, 2-6, 2-7, and 3-2 in Fig.~\ref{fig:ldsrg3_pt4}, where the one- and two-body terms contribute to the non-diagonal part of the Hamiltonian (particle-hole excitations).
As a result, the qDSRG(2) scheme possesses the same asymptotic scaling as the LDSRG(2).
The qDSRG(2) can be easily implement by adding the following recursive contributions to the original LDSRG(2):
\begin{align}
\label{eq:q_dsrg2}
\hat{O}^{k+2}_{\{2\}} &\leftarrow \underbrace{ \frac{k!}{(k+2)!} [[\hat{O}^{k}_{\{2\}}, \hat{A}_2]_3, \hat{A}_{\{2\}}]_{\{2\}} }_\text{LDSRG(3)-1}, \quad k = 0,1,2,\dots .
\end{align}
The qDSRG(2) scheme is analogous to the quadratic commutator approximation described by Neuscamman \textit{et al.}\ in the context of canonical transformation theory.\cite{Neuscamman:2009cy}
However, quadratic canonical transformation theory includes all the diagrams that arise in the double commutator, unlike the qDSRG(2) approach that uses only a smaller subset.

We also investigate an alternative approach, termed LDSRG(2*), based on the LDSRG(3:C2)-1$'$ approximation (imposing $\hat{T}_3 = 0$). In this scheme, we add two extra one-shot corrections to the recursive commutator terms, which are in turn propagated by the linear recursive algorithm to higher-nested commutators.
Thus, the LDSRG(2*) scheme is defined by adding the following to the LDSRG(2) Hamiltonian,
\begin{align}
\hat{O}^{2}_{\{2\}} &\leftarrow \underbrace{ \frac{1}{2} [[\hat{H}, \hat{A}_2]_3, \hat{A}_{\{2\}}]_{\{2\}} }_\text{LDSRG(3;C2)-1$'$} , \label{eq:ildsrg2_2} \\
\hat{O}^{3}_{\{2\}} &\leftarrow \underbrace{ \frac{1}{6} [[[\hat{H}, \hat{A}_{\{2\}}]_{\{3\}}, \hat{A}_{\{2\}}]_3, \hat{A}_{\{2\}}]_{\{2\}} }_\text{LDSRG(3;C2)-1$'$} . \label{eq:ildsrg2_3}
\end{align}
We find that the LDSRG(3;C2)-1$'$ model provides a good approximation to higher order methods such as the LDSRG(3)-2 while requiring fewer additional diagrams.
As shown in Table \ref{table:ildsrg2_stats}, the accuracy of the LDSRG(3;C2)-2 is well reproduced by the LDSRG(3;C2)-1$'$ with small errors ($< 0.01$ m\Eh on average), thus, it justifies the use of the latter approximation in the definition of the LDSRG(2*) scheme.
The closed-form expressions of Eqs.~\eqref{eq:ildsrg2_2} and \eqref{eq:ildsrg2_3} in LDSRG(2*) can be found in the Supplementary Material.
We point out that the LDSRG(2*) theory scales as ${\cal O}(N_{\bf V}^6)$ due to the diagram shown in Fig.~\ref{fig:uccsdt}(b), and therefore, it is impractical for routine use.
However, a comparison with the qDSRG(2) results in Sec.~\ref{sec:results} shows that such expensive terms can be safely neglected without compromising the accuracy.
Therefore, the less expensive qDSRG(2) scheme is the preferred way to add fourth-order terms. 

\begin{table}[h!]
\begin{threeparttable}
\centering
\ifpreprint
\renewcommand{\arraystretch}{1.0}
\else
\renewcommand{\arraystretch}{1.2}
\fi

\caption{Energy deviations (in m\Eh) from the FCI/6-31G values for various LDSRG(3;C2) variants with $\hat{T}_3 = 0$ and $s = 10^3$ \sunit.
Statistics were calculated among twenty-eight closed-shell atoms and molecules listed in Sec.~\ref{sec:results}.
}
\label{table:ildsrg2_stats}

\begin{tabular*}{\columnwidth}{@{\extracolsep{\stretch{1}}} l *{4}{d{1.3}} @{}}
\hline
\hline
Statistics & \multicolumn{1}{c}{-1} & \multicolumn{1}{c}{-1$'$} & \multicolumn{1}{c}{-2} & \multicolumn{1}{c}{CCSD} \\
\hline
Mean signed error\tnote{a} & 4.863 & 4.491 & 4.487 & 4.468 \\
Standard deviation & 5.543 & 5.204 & 5.205 & 5.006 \\
\hline
\hline
\end{tabular*}
\begin{tablenotes}
\footnotesize
\item [a] Mean absolute error is found to be identical as the mean signed error.
\end{tablenotes}
\end{threeparttable}
\end{table}

\subsubsection{DSRG $\Lambda$(T) correction}

In order to formulate a (T) correction for the DSRG, we follow the approach of Kucharski and Bartlett to define the $\Lambda$CCSD(T) method.\cite{Kucharski:1998ie}
First, consider the LDSRG(3) Lagrangian:
\begin{align}
\label{eq:ldsrg3_Lag}
{\cal L}_{\rm LDSRG(3)} =&\, \bar{H}_0
+ \sum_{k=1}^{3} \frac{1}{(k!)^2} \sum_{\vtau_k} \lambda_{\vtau_k} \left( \bar{H}_{\vtau_k} - r_{\vtau_k} \right) ,
\end{align}
where $\bar{H}$ is LDSRG(3) transformed Hamiltonian and $\lambda_{\vtau_k}$ are the Lagrange multipliers for the corresponding amplitude conditions, $\bar{H}_{\vtau_k} - r_{\vtau_k} = 0$.
For brevity, $\vtau_k$ denotes a set of indices with $k$ occupied and $k$ virtual labels.
For example, the complete expression for $\vtau_2$ is
\begin{align}
\sum_{\vtau_2} \lambda_{\vtau_2} \left( \bar{H}_{\vtau_2} - r_{\vtau_2} \right) \equiv \sum_{ij}\sum_{ab} \tens{\lambda}{ij}{ab} \left( \tens{\bar{H}}{ab}{ij} - \tens{r}{ab}{ij} \right).
\end{align}

Next, we isolate the $\hat{A}_3$ contributions to the Lagrangian from those of $\hat{A}_1 + \hat{A}_2$.
Specifically, we write the LDSRG(3) transformed Hamiltonian as a sum of the terms involving \emph{only} $\hat{A}_1$ and $\hat{A}_2$ ($\bar{H}[t_{1,2}]$) plus a remainder ($\bar{H}[t_{1,2,3}]$):
\begin{align}
\bar{H} = \bar{H}[t_{1, 2}] + \bar{H}[t_{1,2,3}].
\end{align}
The $\bar{H}[t_{1,2}]$ term can be further approximated using either qDSRG(2) or LDSRG(2*).
In the following, we shall always take qDSRG(2) as an example, however, the perturbative analysis also holds for LDSRG(2*).
The qDSRG(2) introduces energy errors in the fifth order of perturbation and errors to the cluster amplitudes in the fourth order, so that we can write
\begin{align}
\label{eq:Hbar_t1_t2}
\bar{H}[t_{1,2}] = \bar{H} [{\rm qDSRG(2)}] + {\cal O} (\delta \bar{H}_0^{(5)}, \delta \bar{H}_1^{(4)}, \delta \bar{H}_2^{(4)}),
\end{align}
where $\delta \bar{H}_1^{(4)}$ and $\delta \bar{H}_2^{(4)}$ indicate one- and two-body fourth-order correction terms, respectively.

Now we address perturbative triples corrections on top of qDSRG(2).
Following Kucharski and Bartlett,\cite{Kucharski:1998ie} we assign each quantity a generalized perturbation order (indicated with a superscript surrounded by square brackets), which provides a way to conduct a perturbation theory analysis even to quantities that do not have a well-defined order.
For example, the qDSRG(2) amplitudes are considered to be generalized first-order quantities:
\begin{align}
\hat{A}_{k}^{[1]} &= \hat{A}_{k} [{\rm qDSRG(2)}], \quad k = 1, 2,
\end{align}
as well as the qDSRG(2) Lagrange multipliers
\begin{align}
\lambda^{[1]}_{\vtau_k} &= \lambda_{\vtau_k} [{\rm qDSRG(2)}], \quad k = 1, 2.
\end{align}
The $\hat{A}_3$ amplitudes and $\lambda_{\vtau_3}$ are considered as second-order quantities, which can be easily verified using the $k=3$ term of Eq.~\eqref{eq:ldsrg3_Lag}.
In the following, we choose $\hat{H}^{(0)} \equiv \hat{H}^{(0)}_{\rm Fock}$ as the zeroth-order Hamiltonian in order to obtain a compact and non-iterative set of working equations.
The lowest-order perturbative triples corrections to the qDSRG(2) energy appear in the fourth-order Lagrangian terms of the LDSRG(3) theory:
\begin{align}
\label{eq:E4th_corr_Lag}
{\cal E}^{[4]}_{\Lambda{\rm(T)}} =&\, \bar{H}^{[4]}_0 [t_{1,2,3}] \notag\\
&+ \sum_{k=1}^{2} \frac{1}{(k!)^2} \sum_{\vtau_k} \lambda^{[1]}_{\vtau_k} \bar{H}^{[3]}_{\vtau_k} [t_{1,2,3}] \left( 1 - e^{-s\Delta_{\vtau_k}^2} \right) \notag\\
&+ \frac{1}{36} \sum_{ijk}\sum_{abc} \tens{\lambda}{ijk}{abc,[2]} \left( \tens{\bar{H}}{abc}{ijk,[2]} - \tens{r}{abc}{ijk,[2]} \right).
\end{align}

In Eq.~\eqref{eq:E4th_corr_Lag}, $\bar{H}^{[4]}_0 [t_{1,2,3}]$ is the direct energy contribution due to contractions of $\hat{A}^{[2]}_3$ with $\hat{A}^{[1]}_{1,2}$ and the Hamiltonian:
\begin{align}
\label{eq:E4th_bf}
\bar{H}^{[4]}_0 [t_{1,2,3}] =&\, \frac{1}{2} [[\hat{H}^{(0)}, \hat{A}_3^{[2]}], \hat{A}_3^{[2]}]_0 \notag\\
&+ \frac{1}{2} [[\hat{H}^{(1)}, \hat{A}_2^{[1]}], \hat{A}_3^{[2]}]_0
 + \frac{1}{2} [[\hat{H}^{(1)}, \hat{A}_3^{[2]}], \hat{A}_{1,2}^{[1]}]_0 \notag\\
&+ \frac{1}{6} [[[\hat{H}^{(0)}, \hat{A}_3^{[2]}], \hat{A}_{1,2}^{[1]}], \hat{A}_{1,2}^{[1]}]_0 \notag\\
&+ \frac{1}{6} [[[\hat{H}^{(0)}, \hat{A}_{1,2}^{[1]}], \hat{A}_3^{[2]}], \hat{A}_{1,2}^{[1]}]_0,
\end{align}
where we have excluded the term $\frac{1}{6} [[[\hat{H}^{(0)}, \hat{A}_{1,2}^{[1]}], \hat{A}_{1,2}^{[1]}], \hat{A}_3^{[2]}]_0$, which is null for $\hat{H}^{(0)} = \hat{H}^{(0)}_{\rm Fock}$.
The second term of Eq.~\eqref{eq:E4th_corr_Lag} collects the one- and two-body lambda contributions contracted with third-order one- and two-body DSRG Hamiltonian $\bar{H}_{1,2}^{[3]} [t_{1,2,3}]$, given by
\begin{align}
\label{eq:hbar_3rd}
\bar{H}_{1,2}^{[3]} [t_{1,2,3}] =&\, [\hat{H}^{(1)}, \hat{A}_3^{[2]}]_{1,2} + \frac{1}{2} [[\hat{H}^{(0)}, \hat{A}_{1,2}^{[1]}], \hat{A}_3^{[2]}]_{1,2} \notag\\
&+ \frac{1}{2} [[\hat{H}^{(0)}, \hat{A}_3^{[2]}], \hat{A}_{1,2}^{[1]}]_{1,2}.
\end{align}
The last term of Eq.~\eqref{eq:E4th_corr_Lag} vanishes when the $\hat{A}_3^{[2]}$ amplitudes satisfy the DSRG equation [see Eqs.~\eqref{eq:flow} and \eqref{eq:source}]:
\begin{align}
\label{eq:E4th_b3_flow}
\tens{\bar{H}}{abc}{ijk,[2]} = \left( \tens{\bar{H}}{abc}{ijk,[2]} + \Delta^{ijk}_{abc} \tens{t}{abc}{ijk,[2]} \right) e^{-s (\Delta^{ijk}_{abc})^2}.
\end{align}
Here, the second-order three-body Hamiltonian is given by
\begin{align}
\label{eq:hbar3_2nd}
\bar{H}^{[2]}_3 = [\hat{H}^{(0)}, \hat{A}^{[2]}_3]_3 + [\hat{H}^{(1)}, \hat{A}^{[1]}_2]_3 + \frac{1}{2} [[\hat{H}^{(0)}, \hat{A}^{[1]}_2], \hat{A}^{[1]}_2]_3.
\end{align}

The explicit expressions of $\bar{H}^{[4]}_0 [t_{1,2,3}]$, $\bar{H}_{1,2}^{[3]} [t_{1,2,3}]$, and $\bar{H}^{[2]}_3$ can be found in Appendix \ref{app:expr_hbar_t3}.
At this point, we have specified all terms in Eq.~\eqref{eq:E4th_corr_Lag} and the resulting correction is termed $\Lambda$(T).
The $\Lambda$(T) energy correction requires the knowledge of the qDSRG(2) Lagrange multipliers, which may be obtained by making the qDSRG(2) Lagrangian stationary with respect to all singles and doubles amplitudes.
However, solving for these multipliers exactly is impractical due to the need to recursively evaluate the Hamiltonian and to store an eight-index intermediate.

\subsubsection{DSRG [T] correction}

To formulate a practical perturbative triples correction from the $\Lambda$(T) formalism, we approximate $\lambda_{\vtau_k}^{[1]}$ with the first-order quantities $\lambda_{\vtau_k}^{(1)}$ from a conventional perturbative analysis.
In this regard, we write out the second-order qDSRG(2) Lagrangian:
\begin{align}
\label{eq:pt2_Lag1}
{\cal L}^{[2]}_\text{qDSRG(2)} =&\, [\hat{H}^{(1)}, \hat{A}^{[1]}_{1,2}]_0 + \frac{1}{2} [[\hat{H}^{(0)}, \hat{A}^{[1]}_{1,2}], \hat{A}^{[1]}_{1,2}]_0 \notag\\
&+ \sum_{k=1}^{2} \frac{1}{(k!)^2} \sum_{\vtau_k} \lambda^{(1)}_{\vtau_k} \left( \bar{H}^{[1]}_{\vtau_k} - r^{[1]}_{\vtau_k} \right).
\end{align}
The lambda equations are then given by imposing stationarity of the Lagrangian with respect to variations of singles and doubles amplitudes
\begin{subequations}
\begin{align}
\pdv{{\cal L}^{[2]}_\text{qDSRG(2)}}{\tens{t}{a}{i,[1]}} = 0 \,\,\,\, &\Rightarrow \,\,\,\, \tens{\lambda}{i}{a,(1)} = 2 \left( \frac{ \tens{f}{i}{a} }{ \tens{\Delta}{a}{i} } - \tens{t}{i}{a,[1]} \right), \label{eq:lambda1_approx} \\
\pdv{{\cal L}^{[2]}_\text{qDSRG(2)}}{\tens{t}{ab}{ij,[1]}} = 0 \,\,\,\, &\Rightarrow \,\,\,\, \tens{\lambda}{ij}{ab,(1)} = 2 \left( \frac{ \tens{v}{ij}{ab} }{ \tens{\Delta}{ab}{ij} } - \tens{t}{ij}{ab,[1]} \right). \label{eq:lambda2_approx}
\end{align}
\end{subequations}
Here we have implicitly assumed the use of canonical orbitals that diagonalize the occupied and virtual blocks of the Fock matrix.
Substituting these expressions back to Eq.~\eqref{eq:E4th_corr_Lag} yields a perturbative correction analogous to the [T] method:\cite{Urban:1985bq}
\begin{align}
\label{eq:dsrg[t]}
{\cal E}^{[4]}_{\rm [T]} =&\, \bar{H}^{[4]}_0 [t_{1,2,3}]
 + 2 \sum_{\vtau_1} \bar{H}^{[3]}_{\vtau_1} [t_{1,2,3}] \, {\cal T} (f_{\vtau_1}, t^{[1]}_{\vtau_1}, \Delta_{\vtau_1}) \notag\\
& + \frac{1}{2} \sum_{\vtau_2} \bar{H}^{[3]}_{\vtau_2} [t_{1,2,3}] \, {\cal T} (v_{\vtau_2}, t^{[1]}_{\vtau_2}, \Delta_{\vtau_2}),
\end{align}
where, for brevity, we have introduced the function $\cal T$ :
\begin{align}
{\cal T} (h, t, d) &= h (1 - e^{-sd^2}) / d - t (1 - e^{-sd^2}).
\end{align}

Note that Eq.~\eqref{eq:dsrg[t]} is numerically stable for small energy denominators (e.g., $\Delta_{ab}^{ij} \rightarrow 0$), while the lambda expressions [Eqs.~\eqref{eq:lambda1_approx} and \eqref{eq:lambda2_approx}] are not.
Interestingly, these lambda expressions measure the differences between the qDSRG(2) and MP2 amplitudes.
Numerical tests in Sec.~\ref{sec:results} show that this approximation generally overestimates the magnitude of the qDSRG(2) lambdas.
We also point out that an equivalent way to derive Eq.~\eqref{eq:dsrg[t]} is by considering the fourth-order energy contributions due to the third-order singles and doubles amplitudes from a straightforward order-by-order perturbative analysis of the $\bar{H}_0$ functional based on a Hartree--Fock reference.

\subsubsection{DSRG (T) correction}

The (T) correction considers a Lagrangian from the DSRG second-order perturbation theory (DSRG-PT2):\cite{Wang:2019kf}
\begin{align}
\label{eq:pt2_Lag2}
{\cal L}^{[2]}_\text{DSRG-PT2} =&\, {\cal L}^{[2]}_\text{qDSRG(2)} \left( t^{[1]}_{\vtau_k} \rightarrow t^{(1)}_{\vtau_k} \right), \quad k = 1, 2,
\end{align}
which can be obtained by replacing all qDSRG(2) amplitudes $t^{[1]}_{\vtau_k}$ in Eq.~\eqref{eq:pt2_Lag1} with the DSRG-PT2 counterparts $t^{(1)}_{\vtau_k}$:
\begin{subequations}
\begin{align}
\tens{t}{a}{i,[1]} \rightarrow \tens{t}{a}{i,(1)} &= \tens{f}{i}{a} ( 1 - e^{-s (\Delta_{a}^{i})^2} ) / \Delta_{a}^{i}, \\
\tens{t}{ab}{ij,[1]} \rightarrow \tens{t}{ab}{ij,(1)} &= \tens{v}{ij}{ab} ( 1 - e^{-s (\Delta_{ab}^{ij})^2} ) / \Delta_{ab}^{ij}.
\end{align}
\end{subequations}
The resulting lambda equations are
\begin{align}
\lambda^{(1)}_{\vtau_k} = 2 t^{(1)}_{\vtau_k} e^{-s \Delta^2_{\vtau_k}} / (1 - e^{-s \Delta^2_{\vtau_k}}), \quad k = 1, 2,
\end{align}
which motivates the following approximations
\begin{align}
\lambda^{[1]}_{\vtau_k} \approx 2 t^{[1]}_{\vtau_k} e^{-s \Delta^2_{\vtau_k}} / (1 - e^{-s \Delta^2_{\vtau_k}}), \quad k = 1, 2.
\end{align}
The fourth-order energy correction [Eq.~\eqref{eq:E4th_corr_Lag}] then becomes
\begin{align}
\label{eq:dsrg(t)}
{\cal E}^{[4]}_{\rm (T)} =&\, \bar{H}^{[4]}_0 [t_{1,2,3}]
+ 2 \sum_{k = 1}^{2} \frac{1}{(k!)^2} \sum_{\vtau_k} \bar{H}^{[3]}_{\vtau_k} [t_{1,2,3}] t^{[1]}_{\vtau_k} e^{-s \Delta^2_{\vtau_k}}.
\end{align}

Note that in the limit of $s \rightarrow \infty$, the lambda contributions to Eq.~\eqref{eq:dsrg(t)} vanish. This property is consistent with the fact that in the $s \rightarrow \infty$ limit the DSRG is equivalent to UCC. The latter is a variational approach, and therefore, it can be formulated without Lagrange multipliers.
From this perspective, the (T) correction is consistent with variationality in the limit of $s \rightarrow \infty$, while for finite values of $s$, the contributions from the lambda multipliers counterbalance the reduction in correlation energy due to the non-variational character of the energy.
In contrast, the [T] lambda equations have an incorrect value in the infinite $s$ limit---the lambdas are generally \emph{nonzero} even when the underlying formalism is not truncated (i.e., UCCSD).
In the other limit, $s \rightarrow 0$, it can be easily shown that all DSRG amplitudes are zero and fourth-order corrections for the $\Lambda$(T), [T], and (T) theories are null, which is the correct limit.
The above analysis suggests that,  compared to the [T] correction, the (T) formalism is a better approximation to the $\Lambda$(T) approach because it yields contributions from the lambda amplitudes that are consistent for both boundaries of $s$.

\subsubsection{Comparison of the DSRG $\Lambda$-(T) and $\Lambda$CCSD(T) corrections}

We now briefly compare the DSRG-$\Lambda$(T) energy correction to that of $\Lambda$CCSD(T).
Perhaps the most obvious difference is that the $\Lambda$(T) of DSRG includes a direct energy contribution term $\bar{H}^{[4]}_0 [t_{1,2,3}]$.
The presence of $\bar{H}^{[4]}_0 [t_{1,2,3}]$ is a consequence of $\hat{A}_i$ not commuting with $\hat{A}_k$ for $i \neq k$.
In contrast, in CC theory all components of $\hat{T}$ commute, i.e., $[\hat{T}_i, \hat{T}_k] = 0$.
However, similar terms also arise in the (T) correction of various MRCC formalisms where $[\hat{T}_i, \hat{T}_k] \neq 0$.\cite{Evangelista:2010cq,Hanauer:2012gf}

The other difference between the DSRG-$\Lambda$(T) and $\Lambda$CCSD(T) energy functionals lies in the lambda expressions.
The $\Lambda$CCSD(T) possesses nonzero lambda values because CC theory is not variational.
These lambdas are well approximated using the CCSD cluster amplitudes, adopting which leads to the ``gold-standard'' CCSD(T) energy correction.
On the contrary, the lambdas of DSRG-$\Lambda$(T) are expected to be close to zero in the limit of $s \rightarrow \infty$ as a reflection of the DSRG(2) being variational when no approximations are made in the BCH expansion.
Consequently, for any reasonably large values of $s$, the lambda contributions in the DSRG $\Lambda$(T) or (T) corrections amount to a minor effect compared to the direct term $\bar{H}^{[4]}_0 [t_{1,2,3}]$.

Finally, we point out that the asymptotic scaling of [T] or (T) in DSRG is determined by contractions involving three occupied and four virtual indices.
This ${\cal O}(N_{\bf O}^3 N_{\bf V}^4)$ cost is identical to that of CCSD(T).
In DSRG [T] or (T), the second-order triples amplitudes [Eq.~\eqref{eq:t3_2nd}] differ from those of CCSD(T) by merely an exponential regularizer and they become identical in the limit of $s \rightarrow \infty$.
As such, it is straightforward to implement the DSRG (T) correction by modifying any existing CCSD(T) algorithm.
Another important point is that, like in CCSD(T), an optimal implementation of the DSRG [T] or (T) corrections does not require storage of all triples amplitudes and so may be performed in batches, removing any memory bottleneck.

\section{Results}	
\label{sec:results}

We implemented a proof-of-principle spin-orbital code for all the DSRG approaches in a development branch of \forte,\cite{FORTE2019} using the one- and two-electron integrals from the \PSI quantum chemistry package.\cite{Parrish:2017hg}
The LDSRG(3;C2) variants were tested by directly modifying the LDSRG(3) code, where the recursive evaluation of $\bar{H}_3$ is terminated at the two-nested commutator.
Consequently, the current implementation of LDSRG(3;C2) and its variants do not match the optimal cost in Table \ref{tab:cost}.

\subsection{Atoms and small molecules near equilibrium}
\label{subsec:benchmark}

We benchmarked the DSRG methods on a set of twenty-eight small closed-shell atoms and molecules, including He, Ne, Ar, Be, BH, \ce{BH3}, \ce{Be2}, \ce{BeH2}, \ce{C2}, \ce{C2H2}, \ce{CH2}, \ce{CH2O}, \ce{CH4}, CO, \ce{F2}, \ce{H2}, \ce{H2O}, \ce{H2O2}, HCN, HF, HNC, HNO, HOF, \ce{Li2}, LiH, \ce{N2}, \ce{N2H2}, and \ce{NH3}.
The absolute energies were compared against those of CEPA zero variant (CEPA$_0$),\cite{Meyer:1973fb,Koch:1981dx} CCSD,\cite{PurvisIII:1982kx} CCSD(T),\cite{Raghavachari:1989gf} CC3,\cite{Koch:1997eu} two variants of CCSDT-1 (i.e., CCSDT-1a and CCSDT-1b),\cite{Urban:1985bq} CCSDT,\cite{Noga:1987ea} and FCI.\cite{Sherrill:1999dw}
Unless otherwise notice, the DSRG flow parameter was set to $10^3$ \sunit, a value that is sufficiently large to consider $\tens{\bar{H}}{ij\cdots}{ab\cdots} = 0$ for the molecules addressed here.
All computations were carried out using \PSI, except for the CCSDT-1 and CCSDT results, which were obtained using our in-house spin-orbital code in \forte.
For all computations, we employed Pople's 6-31G basis set\cite{Hehre:1972gg} and core molecular orbitals were excluded from the post-Hartree--Fock treatment of electron correlation, except for BH, \ce{BH3}, Be, \ce{Be2}, \ce{BeH2}, and LiH.
The molecular geometries were directly taken from the experimental data of Computational Chemistry Comparison and Benchmark DataBase\cite{cccbdb} and they are also reported in the Supplementary Material.

\begin{table}[h!]
\begin{threeparttable}
\centering
\ifpreprint
\renewcommand{\arraystretch}{1.0}
\else
\renewcommand{\arraystretch}{1.2}
\fi
\caption{Energy deviations (in m\Eh) of various methods relative to FCI for He, \ce{H2}, and \ce{Li2} where only two electrons are correlated. All DSRG methods use $s = 10^3$ \sunit.}
\label{tab:2e}
\begin{tabular*}{\columnwidth}{@{\extracolsep{\stretch{1}}} l *{3}{d{2.3}} *{3}{d{2.3}} @{}}
\hline
\hline
& \multicolumn{3}{c}{6-31G} & \multicolumn{3}{c}{cc-pVTZ} \\
\cline{2-4} \cline{5-7}
Method & \multicolumn{1}{c}{He} & \multicolumn{1}{c}{\ce{H2}} & \multicolumn{1}{c}{\ce{Li2}} & \multicolumn{1}{c}{He} & \multicolumn{1}{c}{\ce{H2}} & \multicolumn{1}{c}{\ce{Li2}} \\
\hline
CEPA$_0$ & -0.066 & -0.373 & -3.114 & -0.312 & -0.737 & -3.864 \\
LDSRG(2) & -0.133 & -0.637 & -4.155 & -0.396 & -0.980 & -5.418 \\
LDSRG(2*) & 0.003 & 0.005 & 0.114 & 0.001 & 0.004 & 0.078 \\
qDSRG(2)\tnote{a} & 0.000 & -0.000 & 0.003 & 0.000 & -0.000 & -0.003 \\
LDSRG(3)-1$'$ & 0.001 & 0.003 & 0.053 \\
LDSRG(3)-2 & 0.000 & 0.003 & 0.050 \\
LDSRG(3) & 0.000 & 0.004 & 0.056 \\
\hline
\hline
\end{tabular*}
\begin{tablenotes}
\footnotesize
\item [a] Equivalent to LDSRG(3)-1 for two-electron systems.
\end{tablenotes}
\end{threeparttable}
\end{table}

We first present results for correlated two-electron systems, including He, \ce{H2}, and \ce{Li2}.
These systems reveal only the quality of the commutator approximation because triples amplitudes are null.
Table \ref{tab:2e} reports the errors of various methods relative to FCI.
The LDSRG(2) approach strongly overestimates the correlation energies to a similar degree as CEPA$_0$.
The errors are significantly decreased once the induced three-body effects are addressed.
For instance, the \ce{Li2} energy error with the 6-31G basis set is reduced from 4.16 m\Eh [LDSRG(2)] to 0.11 m\Eh [LDSRG(2*)] by improving the BCH expansion, and further reduced to 3 $\mu$\Eh in the qDSRG(2) results.
This improvement is also observed for the larger cc-pVTZ basis set.\cite{Dunning:1989bx,*Prascher:2010eh}
However, the superior accuracy of qDSRG(2) is possibly resulting from error cancellations of the missing diagrams, like in the case of the more sophisticated LDSRG(3) variants.

\begin{table}[th!]
\begin{threeparttable}
\centering
\ifpreprint
\renewcommand{\arraystretch}{0.75}
\else
\renewcommand{\arraystretch}{1.10}
\fi
\caption{Error statistics (in m\Eh) for the twenty-eight molecules computed using various methods comparing against the FCI results. All DSRG computations employ $s = 10^3$ \sunit as the flow parameter. Molecules with the largest error are given in parentheses.}
\label{tab:error_data}
\begin{tabular*}{\columnwidth}{@{\extracolsep{\stretch{1}}}l d{2.3} d{1.3} d{2.3} d{3.3} l @{}}
\hline
\hline
Method & \multicolumn{1}{c}{MSE\tnote{a}} & \multicolumn{1}{c}{MAE\tnote{b}} & \multicolumn{1}{c}{SD\tnote{c}} & \multicolumn{2}{c}{MAX\tnote{d}} \\
\hline
CEPA$_0$ & -3.022 &  4.832 & 14.125 & -71.462 & (\ce{C2}) \\
LDSRG(2)\tnote{e} & -3.471 &  3.617 &  5.029 & -24.744 & (\ce{Be2}) \\
qDSRG(2) &  4.710 &  4.710 &  5.187 &  22.542 & (\ce{C2}) \\
LDSRG(2*) &  4.491 &  4.491 &  5.204 &  24.064 & (\ce{C2}) \\
CCSD &  4.468 &  4.468 &  5.006 &  22.379 & (\ce{C2}) \\

\hline

qDSRG(2)+[T] &  0.576 &  0.973 &  1.315 &  -4.639 & (\ce{C2}) \\
qDSRG(2)+(T) &  0.640 &  0.734 &  0.720 &   1.764 & (\ce{N2}) \\
LDSRG(2*)+[T] &  0.381 &  0.738 &  1.019 &   2.964 & (\ce{CH2}) \\
LDSRG(2*)+(T) &  0.427 &  0.436 &  0.422 &   1.422 & (\ce{Be2}) \\
CCSD(T) &  0.691 &  0.711 &  0.631 &   1.831 & (\ce{N2}) \\

\hline

LDSRG(3;C2)-1 &  1.237 &  1.237 &  1.428 &   6.942 & (\ce{C2}) \\
LDSRG(3;C2)-2 &  0.233 &  0.260 &  0.307 &   1.178 & (\ce{C2}) \\
LDSRG(3)-1 &  0.957 &  0.958 &  0.848 &   2.387 & (\ce{N2}) \\
LDSRG(3)-2 &  0.029 &  0.279 &  0.634 &  -3.002 & (\ce{C2}) \\
LDSRG(3)-1$'$ & -0.257 &  0.326 &  1.025 &  -5.325 & (\ce{C2}) \\
LDSRG(3)-2$'$ &  0.030 &  0.275 &  0.620 &  -2.929 & (\ce{C2}) \\
LDSRG(3) &  0.141 &  0.270 &  0.437 &  -1.654 & (\ce{C2}) \\
CC3 &  0.554 &  0.591 &  0.573 &   1.604 & (\ce{F2}) \\
CCSDT-1a &  0.467 &  0.568 &  0.594 &   1.458 & (\ce{F2}) \\
CCSDT-1b &  0.509 &  0.554 &  0.560 &   1.520 & (\ce{F2}) \\
CCSDT &  0.597 &  0.599 &  0.629 &   1.888 & (\ce{N2}) \\

\hline
\hline
\end{tabular*}
\begin{tablenotes}
\footnotesize
\item [a] Mean signed error (MSE): $\bar{\Delta} = \frac{1}{28} \sum_{i=1}^{28} \Delta_i$, where $\Delta_i = E^{\rm method}_i - E^{\rm FCI}_i$.
\item [b] Mean absolute error (MAE): $\frac{1}{28} \sum_{i=1}^{28} |\Delta_i|$.
\item [c] Standard deviation (SD): $\sqrt{\frac{1}{27} \sum_{i=1}^{28} (\Delta_i - \bar{\Delta})^2}$.
\item [d] Maximum error (MAX): $\Delta_I$, where $I = \mathrm{arg\,max}(|\Delta_i|)$.
\item [e] LDSRG(2) did not converge for \ce{C2} and it was ignored in the statistics.
\end{tablenotes}
\end{threeparttable}
\end{table}

We now compare the error statistics of DSRG methods with other well-established CC theories, as shown in Table \ref{tab:error_data} (the complete data can be found in the Supplementary Material).
In general, all DSRG methods with one- and two-body terms have accuracy comparable to that of CCSD.
Although the LDSRG(2) and CEPA$_0$ results show MAEs similar to that of CCSD, the former fails to converge for \ce{C2}, while the latter shows a very pronounced standard deviation.
The qDSRG(2) (MAE = 4.71, SD = 5.19 m\Eh) and LDSRG(2*) (MAE = 4.49, SD = 5.20 m\Eh), both which include induced three-body terms, reproduce the CCSD results (MAE = 4.47, SD = 5.00 m\Eh) quite well.
In Table~\ref{tab:error_data_ccsd} we show the error statistics for the DSRG methods with one- and two-body terms computed with respect to the CCSD energy.
These data show that adding the fourth-order terms missing from the LDSRG(2) significantly increases the agreement of the qDSRG(2) and LDSRG(2*) methods with the CCSD energy.
Therefore, these two methods can be used as a basis for adding perturbative triples corrections.

\begin{table}[th!]
\begin{threeparttable}
\centering
\ifpreprint
\renewcommand{\arraystretch}{0.75}
\else
\renewcommand{\arraystretch}{1.10}
\fi
\caption{Error statistics (in m\Eh) for the twenty-eight molecules computed using various methods comparing against the CCSD results. All DSRG computations employ $s = 10^3$ \sunit as the flow parameter. Molecules with the largest error are given in parentheses.}
\label{tab:error_data_ccsd}
\begin{tabular*}{\columnwidth}{@{\extracolsep{\stretch{1}}}l d{2.3} d{1.3} d{2.3} d{3.3} l @{}}
\hline
\hline
Method & \multicolumn{1}{c}{MSE\tnote{a}} & \multicolumn{1}{c}{MAE\tnote{a}} & \multicolumn{1}{c}{SD\tnote{a}} & \multicolumn{2}{c}{MAX\tnote{a}} \\
\hline

Excluding \ce{C2}  \\
CEPA$_0$ & -4.292 &  4.334 &  4.654 & -22.433 & (\ce{Be2}) \\
LDSRG(2) & -7.276 &  7.276 &  6.314 & -27.192 & (\ce{Be2}) \\
qDSRG(2) &  0.245 &  0.245 &  0.290 &   0.997 & (\ce{CO}) \\
LDSRG(2*) & -0.039 &  0.137 &  0.188 &  -0.532 & (\ce{N2}) \\

\hline

Including \ce{C2}  \\
CEPA$_0$ & -7.490 &  7.530 & 17.529 & -93.841 & (\ce{C2}) \\
qDSRG(2) &  0.242 &  0.242 &  0.285 &   0.997 & (\ce{CO}) \\
LDSRG(2*) &  0.023 &  0.192 &  0.375 &   1.685 & (\ce{C2}) \\

\hline
\hline
\end{tabular*}
\begin{tablenotes}
\footnotesize
\item [a] Check Table \ref{tab:error_data} for details.
\end{tablenotes}
\end{threeparttable}
\end{table}

We now focus on the results for methods with perturbative triples corrections, which are shown in Table \ref{tab:error_data}.
The overall accuracy measured by a combination of the MAEs and SDs follows the trend: qDSRG(2)+[T] < LDSRG(2*)+[T] < qDSRG(2)+(T) $\sim$ CCSD(T) < LDSRG(2*)+(T).
As discussed in Sec.~\ref{subsec:pt}, the DSRG [T] correction yields undesired nonzero lambda contributions in the limit of $s \rightarrow \infty$.
We would therefore expect that the [T] results should be inferior than those from the DSRG (T) correction, which is indeed what we observe.
For example, the MAE and SD values of qDSRG(2)+(T) are 0.24 and 0.60 m\Eh smaller than those of qDSRG(2)+[T], respectively.
Comparing the two improved LDSRG(2) methods, we find that LDSRG(2*)+[T]/(T) provides statistically more accurate results than those from qDSRG(2)+[T]/(T).
Our analysis shows that in both cases the triples corrections are very close, so that the 0.30 m\Eh MSE difference between the qDSRG(2)+(T) and LDSRG(2*)+(T) is not due to the triples correction, rather, it can be attributed to the higher accuracy of the LDSRG(2*) method, which contains a larger number of fifth-order diagrams [see Fig.~\ref{fig:ldsrg3_pt4.5}].
We also note that for the current benchmark set the error statistics of LDSRG(2*)+(T) (MAE = 0.44, SD = 0.42 m\Eh) are superior to those of the ``gold-standard'' CCSD(T) (MAE = 0.71, SD = 0.63 m\Eh).

Finally, we summarize the results from the iterative triples methods.
The quality of the results (as measured by the MAE and SD) follows the trend: LDSRG(3;C2)-1 < LDSRG(3)-1 < CC3 $\sim$ CCSDT-1a $\sim$ CCSDT-1b $\sim$ CCSDT < LDSRG(3)-1$'$ < LDSRG(3)-2 $\sim$ LDSRG(3)-2$'$ $\lesssim$ LDSRG(3) $\lesssim$ LDSRG(3;C2)-2.
The MAE and SD of LDSRG(3)-1 are 0.96 and 0.85 m\Eh, respectively, and both are notably larger than the corresponding values of CC3 (MAE = 0.59, SD = 0.57 m\Eh) and the CCSDT-1 variants (MAE = 0.55, SD = 0.56 m\Eh for CCSDT-1b).
The error statistics of DSRG are significantly reduced once the fifth-order diagrams are considered, yet increasing the computational scaling to at least ${\cal O}(N_{\bf O}^3 N_{\bf V}^5)$.
For example, the MAE of LDSRG(3)-1$'$ is 0.33 m\Eh, a value that is 0.27 m\Eh smaller than that of CCSDT.
The inclusion of high-scaling terms [${\cal O}(N_{\bf O}^4 N_{\bf V}^5)$ or ${\cal O}(N_{\bf O}^3 N_{\bf V}^6)$] generally provides even more accurate results, as shown by the 0.27 m\Eh MAE and 0.44 m\Eh SD of LDSRG(3).
Interestingly, the most balanced DSRG method with iterative triples appears to be LDSRG(3;C2)-2, yielding an exceptional MAE of 0.26 m\Eh and a minimal SD of 0.31 m\Eh, all at the same cost of CCSDT.
However, further benchmarks are needed to investigate the wave-function quality of the LDSRG(3;C2)-2 theory.

\subsection{Dissociation of \ce{N2}}

In this section, we consider the ground-state potential energy curve (PEC) of \ce{N2} and compare various single-reference CC and DSRG methods against FCI.
We should point out that an accurate description of the entire PEC generally needs multireference generalizations of these single-reference methods.
Nevertheless, this example is useful to assess the robustness of these methods outside of their comfort zone.
Following Ref.~\citenum{Kowalski:2000ee}, we employ Dunning's DZ basis set\cite{DunningJr:1970ho} and freeze the lowest two occupied and highest two virtual orbitals.
The reference wave function is fixed to the determinant $1\sigma_g^2 1\sigma_u^2 2\sigma_g^2 2\sigma_u^2 3\sigma_g^2 1\pi_u^4$ along the dissociation coordinate.

The energy errors with respect to FCI are reported in Table \ref{tab:n2_data} for various multiples of the equilibrium bond length ($r_e = 2.068$ bohr).
Away from the equilibrium bond length, we encounter convergence problems for the LDSRG(2) method, while the improved treatment of the fourth-order terms in the qDSRG(2) and LDSRG(2*) ameliorates the convergence behavior at stretched geometries.
In particular, the least approximate LDSRG(2*) model is found to be numerically robust along the entire PEC.
Similar convergence issues are observed for the LDSRG(3) method and its variants, yet these are less severe than those for LDSRG(2).

\begin{table*}[th!]
\begin{threeparttable}
\centering
\ifpreprint
\renewcommand{\arraystretch}{0.75}
\else
\renewcommand{\arraystretch}{1.10}
\fi
\caption{Errors (in m\Eh) with respect to FCI/DZ (taken from Ref.~\citenum{Kowalski:2000ee}) along the ground-state \ce{N2} potential energy curve. Unless otherwise noted, the flow parameter is set to $s = 10^3$ \sunit. The experimental equilibrium bond length ($r_e$ = 2.068 bohr) is taken from Ref.~\citenum{Kowalski:2000ee}. Points where computations did not convergence are left blank.}
\label{tab:n2_data}
\begin{tabular*}{\textwidth}{@{\extracolsep{\stretch{1}}}l d{1.3} d{2.3} d{3.3} d{3.3} d{3.3} d{4.3} d{4.3} @{}}
\hline
\hline
Method & \multicolumn{1}{c}{$0.75\, r_e$} & \multicolumn{1}{c}{$r_e$} & \multicolumn{1}{c}{$1.25\, r_e$} & \multicolumn{1}{c}{$1.50\, r_e$} & \multicolumn{1}{c}{$1.75\, r_e$} & \multicolumn{1}{c}{$2.00\, r_e$} & \multicolumn{1}{c}{$2.25\, r_e$} \\
\hline

LDSRG(2) & -0.890 &   -3.493 &  -24.773 \\
LDSRG(2) ($s = 1$) &   -0.842 &   -2.177 &    5.951 &   40.694 &  111.045 &  203.236 &  290.041 \\
qDSRG(2) &    3.183 &    8.662 &   20.261 &   29.977 & \\
qDSRG(2) ($s = 1$) &    3.224 &    9.413 &   29.897 &   75.974 &  152.357 &  244.866 &  329.770 \\
LDSRG(2*) &    2.880 &    7.597 &   17.631 &   32.633 &   32.679 &  -14.205 &  -66.836 \\

CCSD\tnote{a} &    3.132 &    8.289 &   19.061 &   33.545 &   17.714 &  -69.917 & -120.836 \\

\hline

qDSRG(2)+[T] &    0.999 &    1.033 &   -4.456 &  -30.248 & \\
qDSRG(2)+(T) &    0.763 &    2.088 &    4.452 &   -2.383 & \\
qDSRG(2)+(T) ($s = 1$) &    0.814 &    2.885 &   14.535 &   47.872 &  111.800 &  195.193 &  274.411 \\
LDSRG(2*)+[T] &    0.697 &   -0.038 &   -7.286 &  -30.458 &  -77.547 & -154.935 & -228.745 \\
LDSRG(2*)+(T) &    0.447 &    0.901 &    1.176 &    0.568 &  -16.272 &  -76.239 & -138.610 \\

CCSD(T)\tnote{a} &    0.742 &    2.156 &    4.971 &    4.880 &  -51.869 & -246.405 & -387.448 \\
CR-CCSD(T)\tnote{a} &    1.078 &    3.452 &    9.230 &   17.509 &   -2.347 &  -86.184 & -133.313 \\

\hline

LDSRG(3;C2)-1 &    0.910 &    2.787 &    7.019 &   11.613 & \\
LDSRG(3;C2)-2 &    0.263 &    0.796 &    2.035 &    4.202 & -6.680 & \\
LDSRG(3)-1 &    0.881 &    2.630 &    6.272 &    5.051 & \\
LDSRG(3)-2 &    0.262 &    0.801 &    1.964 &    1.403 & \\
LDSRG(3)-1$'$ &    0.119 &    0.332 &    0.693 &   -1.114 & \\
LDSRG(3)-2$'$ &    0.261 &    0.792 &    1.920 &    1.408 & \\
LDSRG(3) &    0.286 &    0.960 &    2.682 &    3.315 & -44.595 & \\
LDSRG(3) ($s = 1$) & 0.328 &    1.766 &   13.379 &   49.259 &  118.360 &  207.249 &  290.860 \\

CCSDT-1a &    0.672 &    1.776 &    3.419 & 3.188 &  -24.169 &  -98.749 & -138.682 \\
CCSDT-1b &    0.679 &    1.794 &    3.430 & 3.142 &  -24.339 &  -98.732 & -138.553 \\
CC3 &    0.685 &    1.852 &    3.688 &    3.945 &  -22.619 &  -97.863 & -138.833 \\
CCSDT &    0.580 &    2.107 &    6.064 &   10.158 &  -22.468 & -109.767 & -155.656 \\

\hline
\hline
\end{tabular*}
\begin{tablenotes}
\footnotesize
\item [a] Taken from Ref.~\citenum{Kowalski:2000ee}.
\end{tablenotes}
\end{threeparttable}
\end{table*}

Note that in the absence of approximations of the BCH series, all truncated versions of the DSRG are strictly variational.
Therefore, nonvariational DSRG energies are indicative of the buildup of errors in the DSRG transformed Hamiltonian.
This degradation of the performance of the DSRG is particularly likely to happen when amplitudes are large, which in the case of \ce{N2} is expected for $r > r_e$.
From the data in Table \ref{tab:n2_data}, we note that the general quality of the DSRG methods follows the trend observed in the previous section (Sec.~\ref{subsec:benchmark}).
As expected, the accuracy of these single-reference methods deteriorates as the atomic distance increases.
For instance, the qDSRG(2) error grows from 3.18 m\Eh at $0.75\, r_e$ to 29.98 m\Eh at $1.5\, r_e$, at each point yielding energies that are consistent with those from CCSD.
In contrast, the LDSRG(2) results always fall below the variational minimum and quickly deteriorate for bond lengths greater than $r_e$.
Triples corrections based on the (T) approach are more robust than those based on the [T] scheme, with the latter yielding nonvariational energies already at short bond lengths.
In particular, the qDSRG(2)+(T) results yield errors that are comparable to those of CCSD(T) and show variational behavior up to 1.25 $r_e$, while the LDSRG(2*)+(T) results are similar to those of CR-CCSD(T) at large bond lengths.
In the case of iterative triples, CCSDT shows a quick deterioration of the energy past $r_e$, while the  LDSRG(3) and most of the LDSRG(3)-$n$ approximations appear to be more robust and yield smaller energy errors in the range $[0.75, 1.5]\, r_e$.
Moreover, the quality of approximations that truncate the BCH expansion, decreases at stretched bond lengths as revealed by the LDSRG(3)-$n$ and LDSRG(3;C2)-$n$ results.

Table \ref{tab:n2_data} also reports LDSRG(2) results in which we set the DSRG flow parameter value $s = 1$ \sunit, a typical value employed in multireference versions of this theory.
These results show the effect of energy scale separation in the DSRG, which ultimately results in a suppression of large amplitudes and an improvement of numerical robustness.
In contrast to the unregularized results, all computations with $s=1$ \sunit show convergence across the potential energy curve.
For all regularized DSRG schemes, the energy errors at large bond distances are positive and of the order of 200--300 m\Eh, a result consistent with the fact that suppression of large amplitudes correspond to neglecting static correlation effects, as observed before.\cite{Evangelista:2014kt}

\section{Conclusions}
\label{sec:conclusion}

We have explored a number of approaches to include connected three-body terms (triples) in nonperturbative single-reference unitary many-body theories.
Taking the unitary DSRG formalism as an example, we first investigate the full LDSRG(3) approach that includes single, double, and triple excitations and employs a linear commutator approximation of the BCH expansion in which operators are truncated to three-body terms.
An inspection of the LDSRG(3) terms reveals an ${\cal O}(N_{\bf O}^3 N_{\bf V}^6)$ asymptotic scaling (see Table \ref{tab:l3sdt}), which is identical to that of unitary CCSDT but more expensive than the cost of conventional CCSDT [${\cal O}(N_{\bf O}^3 N_{\bf V}^5)$].

In order to find viable approximations to the LDSRG(3), we perform a perturbative analysis and propose iterative variants,  LDSRG(3)-$n$ ($n = 1,2,3,4$), based on the diagonal Fock zeroth-order Hamiltonian.
The simplest LDSRG(3)-1 method contains only fourth-order diagrams (Fig.~\ref{fig:ldsrg3_pt4}), it scales asymptotically as ${\cal O}(N_{\bf O}^3 N_{\bf V}^4)$, and has a storage requirement of ${\cal O}(N_{\bf O}^2 N_{\bf V}^4)$.
Including additional thirty-two fifth-order diagrams (Figs.~\ref{fig:ldsrg3_pt4.5} and \ref{fig:ldsrg3_pt5}) results in the LDSRG(3)-2 model, with increased computational and storage costs of ${\cal O}(N_{\bf O}^3 N_{\bf V}^5)$ and ${\cal O}(N_{\bf O} N_{\bf V}^5)$, respectively.
These fifth-order diagrams can be split into two batches and one of them requires only ${\cal O}(N_{\bf O}^2 N_{\bf V}^4)$ to store the intermediates (Fig.~\ref{fig:ldsrg3_pt4.5}).
This kind of classification naturally follows from a perturbative analysis based on the zeroth-order Hamiltonian of Fink,\cite{Fink:2006gu} leading to the LDSRG(3)-$n'$ ($n = 1,2,3,4$) theories.
To reduce the storage requirements, we further explore truncating the BCH expansion of the three-body Hamiltonian $\bar{H}_3$ at a finite number of nested commutators.
We find that truncating $\bar{H}_3$ at the two nested commutator reproduces the complete result with sub-m\Eh accuracy.

Several perturbative triples schemes are proposed by identifying the fourth-order energy contributions from triple excitations.
The most general approach based on a Lagrangian formalism, $\Lambda$(T), yields an energy expression [Eq.~\eqref{eq:E4th_corr_Lag}] that contains two components: a direct triples contribution and the lambda contribution.
The former appears due to the non-commuting operators $[\hat{A}_i, \hat{A}_j] \neq 0$ in unitary DSRG, while the latter becomes zero only when the base theory is strictly variational.
To avoid solving the lambda equations, we introduce the [T] and (T) corrections derived by replacing the lambda amplitudes with approximate lambda amplitudes suggested by first-order perturbation theory.
Both the [T] and (T) corrections scale as ${\cal O}(N_{\bf O}^3 N_{\bf V}^4)$ and can be implemented by simply modifying an existing CCSD(T) code.
We note that all these perturbative triples corrections can be applied to other unitary theories without any changes, as long as the singles and doubles amplitudes are determined up to third order in perturbation theory.
Since LDSRG(2) neglects certain fourth-order energy contributions, we have explored two improved approaches: the qDSRG(2) and LDSRG(2*).
The former is equivalent to LDSRG(3)-1 with null triples amplitudes and it preserves the ${\cal O}(N_{\bf O}^2 N_{\bf V}^4)$ computational scaling of LDSRG(2).
The LDSRG(2*) model includes additional 2- and 3-nested commutators found in LDSRG(3)-1$'$ but it is unfeasible in practical computations due to its high computational cost [${\cal O}(N_{\bf V}^6)$].

The DSRG methods are compared to various CC theories on a benchmark set containing twenty-eight closed-shell atoms and small molecules using the 6-31G basis set.
The accuracy of CCSD can be reproduced by both the qDSRG(2) and LDSRG(2*), two methods that include induced three-body intermediates.
Adding the (T) correction to the qDSRG(2) yields results that are as accurate as those from CCSD(T), while the LDSRG(2*)+(T) approach outperforms CCSD(T) with a MAE smaller by 0.28 m\Eh.
For iterative triples methods, the MAE of the simplest LDSRG(3)-1 model is inferior than the CC counterparts with the same computational cost (i.e., CC3 and CCSDT-1) by at least 0.37 m\Eh.
However, the LDSRG(3)-2 results are generally closer to the FCI values than those obtained by CCSDT.
We emphasize that the energies alone do not provide enough information on the overall quality of the wave function and further investigations on molecular properties are therefore desired.

This work paves a way forward to addressing perturbative triples in the MR-DSRG formalism.
In particular, the (T) correction of Eq.~\eqref{eq:dsrg(t)} can be easily generalized to the MR case.
However, questions remain on how to introduce the three-body diagrams\cite{Zhang:2019ec} that are missing in the MR-LDSRG(2) and are necessary to create a balanced perturbative triples energy.
One promising route is to develop the MR extension of the qDSRG(2) ansatz, since its computational cost is identical to that of LDSRG(2).
If the accuracy achieved by these SR-DSRG methods could be reproduced in their multireference counterparts, we anticipate that one could formulate a useful  MR-DSRG scheme with perturbative triples to quantitatively predict the energy and properties of strongly correlated systems.

\section*{Supplementary Material}

See the supplementary material for
1) the higher-order diagrams of LDSRG(3),
2) the closed-form expressions added in LDSRG(2*) and qDSRG(2),
3) the equilibrium geometries of the twenty-four molecules taken from the Computational Chemistry Comparison and Benchmark DataBase,
4) the complete energetics of the twenty-eight molecules and atoms computed using various DSRG and CC methods,
and 5) the potential energy curve of \ce{N2} obtained using various DSRG methods with $s = 1.0$ \sunit.

\section*{Data Availability Statement}
The data that supports the findings of this study are available within the article and its supplementary material.

\begin{acknowledgments}
C.L. and F.A.E. were supported by the U.S. Department of Energy under Award No. DE-SC0016004, a Research Fellowship of the Alfred P. Sloan Foundation, and a Camille Dreyfus Teacher-Scholar Award.
\end{acknowledgments}

\titleformat{\section}[block]
  {\normalfont\sffamily\bfseries}
  {\MakeUppercase{\appendixname~\thesection.}}{0.5 em}{\uppercase}
\titlespacing{\section}{0pt}{12pt}{12pt}

\appendix

\section{Expressions of the DSRG $\Lambda$(T) correction}
\label{app:expr_hbar_t3}

In this appendix, we report the explicit expressions for Eqs.~\eqref{eq:hbar3_2nd}, \eqref{eq:E4th_bf} and \eqref{eq:hbar_3rd}.
For brevity, Einstein summation convention over repeated indices is assumed in the following.
We first express the second-order three-body Hamiltonian given by Eq.~\eqref{eq:hbar3_2nd}.
Note that we are only interested in those components appearing in the triples amplitudes equation [Eq.~\eqref{eq:E4th_b3_flow}].
To this end, the last term of Eq.~\eqref{eq:hbar3_2nd} $\frac{1}{2} [[\hat{H}^{(0)}, \hat{A}^{[1]}_2], \hat{A}^{[1]}_2]_3$ does not contribute to $\tens{\bar{H}}{abc}{ijk,[2]}$ for $\hat{H}^{(0)} = \hat{H}^{(0)}_{\rm Fock}$ and it can be ignored here.
We then obtain
\begin{align}
\tens{\bar{H}}{abc}{ijk,[2]} =&\, \permop (a/bc) \tens{f}{a}{d} \tens{t}{bcd}{ijk,[2]} - \permop (i/jk) \tens{f}{l}{i} \tens{t}{abc}{jkl,[2]} + \tens{w}{abc}{ijk,[2]},\\
\tens{w}{abc}{ijk,[2]} =&\, \permop (k/ij) \permop (a/bc) \tens{v}{al}{ij} \tens{t}{bc}{kl,[1]} + \permop (i/jk) \permop (c/ab) \tens{v}{ab}{di} \tens{t}{cd}{jk,[1]},
\end{align}
where the index permutation operator is defined as $\permop (p/rs) f(p,r,s) = f(p,r,s) - f(r,p,s) - f(s,r,p)$ for a quantity $f(p,r,s)$ labeled by indices $p$, $r$, and $s$.
In the canonical basis, the first two terms of $\tens{\bar{H}}{abc}{ijk,[2]}$ can be simplified as $\Delta_{ijk}^{abc} \tens{t}{abc}{ijk,[2]}$ and thus the second-order triples amplitudes are compactly expressed as
\begin{align}
\label{eq:t3_2nd}
\tens{t}{abc}{ijk,[2]} =&\, \tens{w}{abc}{ijk,[2]} \big[ 1 - e^{-s (\Delta^{ijk}_{abc})^2} \big] \big/ \Delta^{ijk}_{abc}.
\end{align}

The direct energy contribution of triples $\bar{H}^{[4]}_0 [t_{1,2,3}]$ is given by
\begin{widetext}
\begin{align}
\bar{H}^{[4]}_0 [t_{1,2,3}] =&\, 
 \underbrace{ \frac{1}{4} (f^{ i }_{ a } t^{ j k }_{ b c } t^{ i j k }_{ a b c } + v^{ i j }_{ a b } t^{ k }_{ c } t^{ i j k }_{ a b c } ) 
- \frac{1}{2} ( v^{ c i }_{ a b } t^{ j k }_{ c d } t^{ i j k }_{ a b d } + v^{ j k }_{ a i } t^{ i l }_{ b c } t^{ j k l }_{ a b c } )}_{\frac{1}{2} [[\hat{H}^{(1)}, \hat{A}_2^{[1]}], \hat{A}_3^{[2]}]_0 + \frac{1}{2} [[\hat{H}^{(1)}, \hat{A}_3^{[2]}], \hat{A}_{1,2}^{[1]}]_0}
+ \underbrace{\frac{1}{12} ( f^{ b }_{ a } t^{ i j k }_{ a c d } t^{ i j k }_{ b c d } - f^{ j }_{ i } t^{ i k l }_{ a b c } t^{ j k l }_{ a b c } )}_{\frac{1}{2} [[\hat{H}^{(0)}, \hat{A}_3^{[2]}], \hat{A}_3^{[2]}]_0} \notag\\
&+
\underbrace{\frac{1}{2} ( f^{ j }_{ i } t^{ k }_{ a } t^{ i l }_{ b c } t^{ j k l }_{ a b c } - f^{ b }_{ a } t^{ i }_{ c } t^{ j k }_{ a d } t^{ i j k }_{ b c d } ) 
+ \frac{1}{4} ( f^{ b }_{ a } t^{ i }_{ a } t^{ j k }_{ c d } t^{ i j k }_{ b c d } - f^{ j }_{ i } t^{ i }_{ a } t^{ k l }_{ b c } t^{ j k l }_{ a b c } )}_{\frac{1}{6} [[[\hat{H}^{(0)}, \hat{A}_3^{[2]}], \hat{A}_{1,2}^{[1]}], \hat{A}_{1,2}^{[1]}]_0 + \frac{1}{6} [[[\hat{H}^{(0)}, \hat{A}_{1,2}^{[1]}], \hat{A}_3^{[2]}], \hat{A}_{1,2}^{[1]}]_0} .
\end{align}
\end{widetext}
The one- and two-body third-order transformed Hamiltonian due to triples are evaluated as
\begin{align}
\tens{\bar{H}}{a}{i,[3]} [t_{1,2,3}] =&\, \frac{1}{4} v^{ j k }_{ b c } t^{ i j k }_{ a b c }
+ \frac{1}{8} ( f^{ b }_{ a } t^{ j k }_{ c d } t^{ i j k }_{ b c d } - f^{ j }_{ i } t^{ k l }_{ b c } t^{ j k l }_{ a b c } ) \notag\\
&+ \frac{1}{2} ( f^{ c }_{ b } t^{ j k }_{ b d } t^{ i j k }_{ a c d } - f^{ k }_{ j } t^{ j l }_{ b c } t^{ i k l }_{ a b c } ), \\
\tens{\bar{H}}{ab}{ij,[3]} [t_{1,2,3}] =&\, f^{ k }_{ c } t^{ i j k }_{ a b c } + f^{ d }_{ c } t^{ k }_{ c } t^{ i j k }_{ a b d } - f^{ l }_{ k } t^{ k }_{ c } t^{ i j l }_{ a b c } \notag\\
&+ \frac{1}{2} [ \permop ( i / j ) f^{ k }_{ i } t^{ l }_{ c } t^{ j k l }_{ a b c } - \permop ( a / b ) f^{ c }_{ a } t^{ k }_{ d } t^{ i j k }_{ b c d } ] \notag\\
&- \frac{1}{2} [ \permop ( i / j ) v^{ k l }_{ c i } t^{ j k l }_{ a b c } + \permop ( a / b ) v^{ c d }_{ a k } t^{ i j k }_{ b c d } ],
\end{align}
where the index permutation operator is defined as $\permop (p/q) f(p,q) = f(p,q) - f(q,p)$.
Note that we do not assume canonical orbitals in these expressions.


%

\end{document}